\documentclass[aps,prl,twocolumn,superscriptaddress]{revtex4-2}
\usepackage{tikz}
\usepackage[english]{babel}
\usepackage[utf8]{inputenc}
\usepackage{indentfirst}
\usepackage{amsmath}
\usepackage{amssymb}
\usepackage{eufrak}
\usepackage{graphicx}
\usepackage{psfrag}

\newcommand{\Lax}{A}

\newcommand{\La}{\mathcal{L}}

\newcommand{\ZZ}{\mathcal{Z}}

\newcommand{\valos}{\mathbb{R}}

\usepackage{amsthm}

\newcommand{\vev}[1]{\left\langle #1 \right\rangle}
\newcommand{\ket}[1]{{\left|#1\right\rangle}}
\newcommand{\bra}[1]{{\left\langle #1\right|}}

\newcommand{\rhogamal}{\rho_\gamma(\alpha)}

\newcommand{\secc}[1]{\textit{#1}.---}

\usepackage{ifpdf}

\ifpdf
\usepackage{epstopdf}
\usepackage[pdftex,colorlinks,urlcolor=blue,citecolor=blue,linkcolor=blue]{hyperref}
\else
\usepackage[hypertex,colorlinks,urlcolor=blue,citecolor=blue,linkcolor=blue]{hyperref}
\fi
\begin{document}

\title{
  Hidden quasi-local charges and Gibbs ensemble in a Lindblad system
}
\author{Marius de Leeuw}
\affiliation{School of Mathematics
\& Hamilton Mathematics Institute, 
Trinity College Dublin, Ireland}
\author{Chiara Paletta}
\affiliation{School of Mathematics
\& Hamilton Mathematics Institute, 
Trinity College Dublin, Ireland}
\author{Bal\'azs Pozsgay}
\affiliation{MTA-ELTE “Momentum” Integrable Quantum Dynamics Research Group, Department of Theoretical Physics, 
E\"otv\"os Lor\'and University, Budapest, Hungary}
\author{Eric Vernier}
\affiliation{Laboratoire de Probabilit\'es, Statistique et Mod\'elisation
CNRS - Univ. Paris Cit\'e - Sorbonne Univ.
Paris, France}

\begin{abstract}
We consider spin-1/2 chains with external driving that breaks the continuous symmetries of the Hamiltonian. We introduce
a family of models described by the Lindblad equation with local jump operators. The models have hidden strong
symmetries in the form of quasi-local charges, leading to multiple non-equilibrium steady states. We compute them
exactly in the form of Matrix Product Operators, and argue that they are the analogues of quantum many body scars in the
Lindbladian setting. We observe that the dynamics leads to the emergence of a Gibbs ensemble constructed from the hidden
charges.  
\end{abstract}

\maketitle

\secc{Introduction}
If a small physical system is made into contact with a much larger system  (the bath), which is itself in thermal equilibrium, 
then the interaction with the bath will typically induce thermalization of the small system: in
the long time limit all details of its initial state will be washed away and 
its emerging steady state will be determined by the thermodynamical state functions of the bath
\cite{necessity-of-eth}. This is a general phenomenon in both the classical and in the quantum world, and it
is essential for the formulation of statistical physics and thermodynamics.

A similar phenomenon happens also in
situations with external driving \cite{pumping,ness-cite1}. Typically there is a unique steady state, whose properties
depend only on the 
parameters of the driving, and all properties of the initial states are eventually lost during time evolution. Quantum
many-body systems with driving (or simply in contact with their environment) can often be described by the Lindblad
equation \cite{lindblad-intro}, and generic Lindblad systems have a unique non-equilibrium steady state (NESS)
\cite{lindblad-unique-1}.

Models with a non-unique NESS are exceptional: they conserve additional information about the initial state
\cite{multiness-phd-thesis}. They are analogous to isolated systems with ergodicity breaking, which 
 have been well studied in the last two decades. Today various mechanisms leading to ergodicity breaking are known 
\cite{huse-review,essler-fagotti-quench-review,hfrag-review}, and all of them are associated with exotic
symmetries of the system.

We focus on the question: What are possible ways to have multiple NESS
in a many-body Lindblad system? Similar to ergodicity breaking, non-uniqueness of the NESS is associated with
the presence of extra conservation laws.
In Lindblad systems conserved quantities can be constructed if the model has
so-called strong symmetries 
\cite{prosen-strong-symmetries,lindblad-symmetries,enej-prosen-spin1degness,buca-strong-symmetries}.

In this work we uncover a novel mechanism leading to  unexpected degenerate NESS in
a Lindblad system. 
We introduce a model with a
local Hamiltonian and local jump operators in the bulk, which break the standard $U(1)$ symmetry of the
Hamiltonian. Nevertheless we find hidden strong symmetries in the form of quasi-local charges: extensive operators with
a quasi-local operator 
density. Previously such operators were treated in the context of the Generalized Gibbs Ensemble
\cite{prosen-enej-quasi-local-review,JS-CGGE}, and  our
work is the first one to uncover quasi-local charges in a Lindblad system with local driving in the bulk.

We also find explicit and exact formulas for the degenerate NESS in our model: we present them as Matrix Product
Operators (MPO) with fixed bond dimension.
We argue that they are analogous to the quantum many body scars known from Hermitian systems \cite{hfrag-review,fragmentation-scars-review-2}.
We also consider time evolution from selected initial states, and rigorously compute the steady state values of selected
observables, thereby proving that the system retains memory of the initial state. Furthermore, we show that in the
infinite volume limit the emerging steady states can also be described by a Gibbs Ensemble constructed from the hidden
quasi-local charge.

\secc{Lindblad systems} We consider the dynamics of a quantum spin-1/2 chain in contact with its environment. If the
environment is Markovian, the time evolution of the density matrix $\rho$ of the system can be described by
the Lindblad equation, which reads

\begin{equation}
 \dot\rho =i[\rho,H]+\sum_a u_a \Big[
    \ell_a\rho \ell_a^\dagger-\frac{1}{2}\{\ell^\dagger_a \ell_a,\rho\}\Big],
    \label{lindbeq}
\end{equation}
and equivalently in the superoperator formalism  
$\dot\rho={\mathcal{L}}\rho$,  
where ${\mathcal{L}}$ is the so-called Lindblad superoperator \cite{lindblad-eredeti,breuer2002theory}. 

Here $H$ is the Hamiltonian of the system and $\ell_a$ are the jump operators, which describe
processes mediated by the environment. The parameters $u_a\in \valos^+$ are coupling constants,
and the index $a$ labels the various jump operators.

We are interested in models where the jump operators are localized in real space, and the system is
translationally invariant. Furthermore, we consider  periodic boundary conditions and one
family of jump operators in the bulk. In such a case $u_a\equiv U$ with a uniform coupling $U$ and $\ell_a\equiv
\ell(j)$ is a 
fixed short range operator localized around the site $j$.

\secc{Symmetries and NESS} In a Lindblad system the non-equilibrium steady states (NESS) are the density matrices $\rho$
which emerge in the long time limit, and they satisfy  $\La \rho=0$.
In a generic Lindblad system without symmetries there is a unique NESS, but counterexamples are also known
\cite{lindblad-unique-1,enej-prosen-spin1degness}. In such exceptional cases the system
preserves memory of the initial 
state, because different initial density matrices 
evolve to different NESS in the long time limit.
One of the possible ways to have non-unique NESS is to have conservation laws in the model, because different initial
mean values 
of the conserved quantity necessarily lead to multiple NESS. 

Conservation laws are typically associated with symmetries.
In Hermitian quantum mechanics, symmetries are represented by linear operators which commute with the
Hamiltonian, and every symmetry automatically leads to a conservation law for an observable quantity.
The situation is very different in the non-Hermitian
setting of the Lindblad equation \cite{lindblad-symmetries}. In these systems a symmetry operation might or might not
lead to a conserved quantity, and not all conserved quantities originate in symmetries.

However, there is a direct connection in the case of  a ``strong symmetry''. We
say that an operator $Q$ is a strong symmetry of a Lindblad system, if $Q$ commutes with the Hamiltonian $H$ and all
jump operators individually. In this case $\La^\dagger Q=\La Q=0$ and thus $Q$ is also a NESS.
Of special interest are those strong symmetries which are represented by {\it extensive operators}, ie.
$Q=\sum_j q(j)$, where $q(j)$ is the operator density of the conserved charge. 

\secc{The Hubbard Lindbladian} An example for a Lindblad system with such a strong symmetry was considered in
\cite{essler-prosen-lindblad}. Using the notation $X_j, Y_j, Z_j$ for the Pauli matrices
acting on site $j$ of the spin chain, we can write the Hamiltonian and the jump operators of the model of
\cite{essler-prosen-lindblad} as
\begin{equation}
  \label{proessl}
  H=\sum_j X_jX_{j+1}+Y_{j}Y_{j+1},\qquad \ell(j)=Z_j.
\end{equation}
The system is homogeneous with a global coupling constant $U$.

Here the Hamiltonian describes the so-called XX model, while the jump operators
describe local de-phasing effects. Substituting \eqref{proessl} into \eqref{lindbeq}, the resulting Lindblad superoperator can be seen as the Hubbard model with imaginary
coupling constant \cite{essler-prosen-lindblad}, which implies that the superoperator is Yang-Baxter integrable, and the
Lindblad superoperator can be diagonalized using Bethe Ansatz.

This model has an extensive strong symmetry given by 
\begin{equation}
  Q_0=\sum_j Z_j,
\end{equation}
which is the global magnetization. Accordingly, in this model the NESS is not unique and in a finite volume $L$ the null space of the superoperator $\La$ is $L+1$ dimensional. Representative NESS can be chosen as the
$L+1$ projectors $P_N$ to the different sectors of the Hilbert space with a given total magnetization $N$. Alternatively, an
over-complete basis for the null-space can be chosen as
\begin{equation}
\rho(\alpha)\sim  e^{\alpha Q_0}=\prod_j e^{\alpha Z_j}, \qquad \alpha\in\valos .
\end{equation}
These density matrices are linear combinations of $P_N$. They are product operators in real space: their operator space
entanglement is zero.

\secc{Our model} We consider a deformation of the model given by \eqref{proessl}. In our case, the Hamiltonian is 
\begin{equation}
  \label{HXY}
  H=\sum_j X_jY_{j+1}-Y_{j}X_{j+1},
\end{equation}
which is known as the Dzyaloshinskii–Moriya interaction term. It can be related to the XX Hamiltonian \eqref{proessl}
by applying an homogeneous twist along the chain \cite{ericZn}. 
We have a global coupling constant $U$, and the jump operators are given by
\begin{multline}
  \label{ljump}
    \ell(j)=\frac{1}{1+\gamma^2}\left(Z_{j+1}+\gamma (X_j +X_{j+2})X_{j+1}\right.\\\left.
      -\gamma^2 X_j Z_{j+1}X_{j+2}\right),
 \end{multline}
where $\gamma\in\valos$ is seen as a
deformation parameter, such that $\gamma=0$ describes the original model \eqref{proessl}. 
The jump operator \eqref{ljump} acts non-trivially on three neighbouring sites and satisfies the special relations
 \begin{equation}
   \label{ellsq}
(\ell(j))^\dagger=\ell(j),\qquad   (\ell(j))^2=1.
 \end{equation}
 Neighbouring jump operators do
not commute, but $[\ell(j),\ell(k)]=0$ if $|j-k|\ge 2$.

For simplicity we consider the regime $0<\gamma<1$ in all of this paper.  Other regimes can be treated by special
similarity and duality transformations.  Furthermore, the points $\gamma=\pm 1$ require special care due to
extra $U(1)$ charges, which enlarge the null space of the Lindbladian. The other regimes and the special points deserve
a separate study. 

The model can also be formulated in terms of fermion operators, following the usual Jordan-Wigner transformation \cite{XX-original}. Introducing the Majorana operators $\psi_{2j-1}=X_j \prod_{l<j} Z_l$, $\psi_{2j}=Y_j \prod_{l<j} Z_l$, which satisfy $\{ \psi_a,\psi_b \} = 2\delta_{a,b}$, we have 
\begin{equation} 
H = \sum_{k} \psi_{k-1} \psi_{k+1} \,,  
\label{eq:Hfermions}
\end{equation} 
where the sum is now over twice the number of sites of the original spin model. Considering the spin chain defined on
$L$ sites with periodic boundary conditions translates in the Majorana language into $\psi_{L+k}=\mathcal{Z} \psi_{k}$,
where $\mathcal{Z} \equiv (-1)^F\equiv \prod_j Z_j$ is the fermion number parity.
The jump operators take the form 
\begin{equation}
\ell(j) = 
\frac{i}{1+\gamma^2} \left( \psi_{2j+2} - \gamma \psi_{2j}  \right) \left( \psi_{2j+1} - \gamma \psi_{2j+3}  \right) \,.  
\label{eq:lfermions}
\end{equation}

The jump operators break the $U(1)$ symmetry of the original model: they induce particle creation and
annihilation, but due to conservation of $\ZZ$ creation an annihilation
happens in pairs. 

While the Hamiltonian \eqref{eq:Hfermions} is bilinear in terms of the Majorana operators and can therefore be diagonalized using free-fermion techniques \cite{XX-original}, the jump operators \eqref{eq:lfermions} introduce quartic terms in the Lindblad equation \eqref{lindbeq}, and our model is therefore truly interacting.

\secc{Integrability}
 The work \cite{essler-prosen-lindblad} initiated the study of integrable Lindbladians: these are
models where the superoperator originates from solutions of the Yang-Baxter equation. Recently a systematic search was
initiated to find integrable Lindbladians \cite{sajat-lindblad} (see also \cite{essler-lindblad-review}), and the present model was discovered with the same
methods.
The model given by \eqref{proessl} can be related to the Hubbard model, whereas our Lindblad superoperator is related to
the deformation of the Hubbard model treated in the recent work \cite{sajat-hubbard}.
Therefore our model is also Yang-Baxter integrable.
Interestingly, the derivations below do  not make use of this property. They will, however, make use of the ``superintegrability'' property of the Hamiltonian \eqref{HXY}, namely the fact that it allows for non-abelian families of conserved charges, which commute with $H$ but not necessarily with one another \cite{superintegrability,ericZn} (see \cite{Note1} for a detailed discussion).
 
\secc{Main results} We find that our Lindbladian  possesses a null space which is $L+1$ dimensional in a
finite volume $L$. The existence of the degenerate NESS is explained by an unexpected strong symmetry in
the system.
 This symmetry and the associated conserved charge are obtained from the original $Q_0$ of the un-deformed
model via a non-local transformation, which is performed by a Matrix Product Operator (MPO).

More specifically, let us define the MPO $T(\gamma)$ as
\begin{equation}
T(\gamma) = \mathrm{Tr}_{\mathcal{A}} ( \Lax_{L}(\gamma) \Lax_{L-1}(\gamma)  \ldots  \Lax_1(\gamma) ) \,.
\end{equation}
Here $\mathcal{A}$ is a two-dimensional ancillary space, and the tensor $\Lax(\gamma)$ is written with respect to this
space as
\begin{equation}
\Lax_j(\gamma)
 = \frac{1}{2}
\left(  
\begin{array}{cc}
g^- + g^+Z_j & g^+ X_j -  i g^- Y_j \\
g^- X_j +  i g^+Y_j &   g^+ - g^-Z_j \\ 
\end{array}
 \right)  \,,
\end{equation}
where $g^\pm=\sqrt{1\pm \gamma}$. The operators $T(\gamma)$ form a mutually commuting family, namely $[T(\gamma),T(\gamma')]=0$: in \cite{Note1} we show that they can be recast as a series expansion in powers of $\gamma$, whose coefficients are expressed in terms of the mutually commuting conserved charges of $H$. From there, we further show that the operators
$T(\gamma)$ and $T(\gamma)^\dagger$ obey the property: 
\begin{equation}
T(\gamma)T(\gamma)^{\dagger}  = T(\gamma)^{\dagger} T(\gamma)  = 1 + \gamma^L \mathcal{Z} \,. 
\label{eq:TTdag}
\end{equation}
Hence, in the $L\to \infty$ limit they become inverse of each other.  

Next, we define the deformation of $Q_0$ as
\begin{equation} \label{Qgammadef}
  Q_\gamma= T(\gamma)^\dagger Q_0 T(\gamma),
\end{equation}
which in the $L \to \infty$ limit corresponds to a conjugation relation.
This conjugation can be understood as a quasi-local deformation of $Q_0$,
involving the non-abelian conserved charges of the Hamiltonian \eqref{HXY}. $Q_\gamma$ remains an extensive
operator, but its operator density $q_\gamma(j)=T(\gamma)^{\dagger} Z_j T(\gamma)$ becomes quasi-local; details are
given in  \footnote{Supplemental  Materials to ``Hidden quasi-local charges and Gibbs ensemble in a Lindblad system''}.

In \cite{Note1} we show that the operator $Q_\gamma$ is a strong symmetry of the Lindbladian: it commutes with the Hamiltonian
\eqref{HXY} 
and also with the jump operators \eqref{ljump}. This implies that it is a conserved charge for the Lindbladian time evolution.

We further find that the matrices
\begin{equation}
  \label{rhodef}
  \rhogamal=T(\gamma)^\dagger e^{\alpha Q_0} T(\gamma)=
  T(\gamma)^\dagger \left[\prod_j e^{\alpha Z_j}\right] T(\gamma)
\end{equation}
are (un-normalized) density matrices: they are Hermitian and positive definite. 
They are also strong symmetries.
It follows, that the matrices $\rho_\gamma(\alpha)$, $\alpha\in\valos$  
 are NESS of the Lindbladian with fixed deformation parameter $\gamma$ and
arbitrary coupling strength $U$. Alternatively, we could consider the density matrices  $\tilde{\rho}_\gamma(\alpha) =
T(\gamma)^{-1} e^{\alpha Q_0} T(\gamma)$, which coincide with \eqref{rhodef} up to corrections of order $\gamma^L$. 

The operators $\rho_\gamma(\alpha)$, $\alpha\in\valos$   form an overcomplete basis for the null space
of the Lindbladian, which has dimension $L+1$ in a finite volume $L$. This can be proven by expanding
$\rho_\gamma(\alpha)$ into a power series in $\alpha$: this produces the powers of $Q_\gamma$ (up to corrections of the
order $\gamma^L$), which (together with the identity) span a space of dimension $L+1$.

Steady states in MPO form have been found earlier in multiple instances in the literature (for systems with boundary
driving see for example 
\cite{prosen-boundary-lindblad-1,enej-prosen-spin1degness,prosen-exterior-lindblad,prosen-boundary-lindblad-2}). Our
results are unique because 
we treat a system locally driven in the bulk, and the bond dimension of the MPO is a fixed small number.

\secc{Frustration free property and Lindbladian scars}
The density matrices $\rhogamal$ can be written as an MPO with bond dimension 4. Therefore, their operator space
entanglement satisfies an area 
law. Interestingly, the $\rhogamal$ are related to frustration free Hamiltonians.

To see this, we define an auxiliary Hermitian superoperator $M$, which acts on any $\rho$ as
\begin{equation}
  M\rho=\sum_j  \ell(j)  \rho \ell^\dagger(j).
\end{equation}
In our case, the strong symmetry and the relations \eqref{ellsq} imply that
$\rhogamal$ are eigenvectors of $M$ with eigenvalue $L$, and that is the maximal possible
eigenvalue of $M$. By definition, this means that the superoperator $M$ is frustration free.

A related model with the frustration free property was investigated in \cite{MPS-cluster-model} (see also \cite{peschel-emery,frustration-witten}). Their Hamiltonian acts
on the spin-1/2 Hilbert space and it can
be written as
\begin{equation}
  K=\sum_j \ell(j).
  \label{FFK}
\end{equation}
It has two extremal states $\ket{\Psi_\pm}$ satisfying the frustration free condition $\ell(j)\ket{\Psi_\pm}=\pm
\ket{\Psi_\pm}$. It follows that the density matrices $\rho_\pm=\ket{\Psi_\pm}\bra{\Psi_\pm}$ are frustration free
eigenstates of $M$. Furthermore, they are NESS for our Lindbladian, and they are reproduced by $\rhogamal$ in the
$\alpha\to \pm\infty$ limit. Our procedure to obtain the density matrices $\rhogamal$ can be seen as a generalization of
the methods of \cite{MPS-cluster-model} to the Lindbladian setting.

After re-normalization and shifting by a matrix proportional to the identity, the action of the full superoperator can be written as
\begin{equation}
\tilde\La\rho\equiv(U^{-1}\La+L)\rho=M\rho+i\, U^{-1}[\rho,H].
\end{equation}

The superoperator $\tilde\La$ becomes Hermitian \footnote{We would like to clarify that we are referring about the Hermiticity property of the action of the superoperator $\tilde{\mathcal{L}}$. This is not the same as Hermiticity or anti-Hermiticity of the commutator $[\rho,H]$.} for $U=iu$, $u\in\valos$. In such a case $\rhogamal$ are still
eigenoperators of $\tilde\La$, they have low spatial entanglement, and they are in the middle of the spectrum for a
generic real $u$. Therefore, they can be seen as quantum many body scars of $\tilde\La$
\cite{hfrag-review,fragmentation-scars-review-2}. We suggest to call them {\it Lindbladian scars} for our original superoperator
$\La$ \footnote{Closely related notions of a Lindbladian scar appeared in \cite{buca-lindblad-scar,lindblad-scar-embed},
  see also \cite{lindblad-scar-szeru}.}.  

\secc{Mean values} The physical properties of $\rhogamal$ can be demonstrated by computing the mean
values of local observables in these states, which can be done using standard MPO techniques \cite{Note1}. First we compute the
mean value of the local operator $Z$ placed at any site $j$. We find
\begin{equation}
\vev{Z_j}=\frac{\text{Tr}(\rhogamal Z_j)}{\text{Tr}\rhogamal } =  \frac{ \left(1-\gamma ^2\right)  \tanh(\alpha) }
  { (\gamma  \tanh(\alpha))^L+1}
  \label{Zjmeanv}
\end{equation}
In the large volume limit this gives
\begin{equation}
\vev{Z_j}_{L\to\infty}= \left(1-\gamma ^2\right)  \tanh(\alpha) .
\end{equation}
In the un-deformed model ($\gamma=0$) the mean value is $\tanh(\alpha)$, thus the transformation
\eqref{rhodef} decreases the mean value by a factor that depends only on $\gamma$.

It is also useful to consider a measure for the breaking of the standard $U(1)$-symmetry.
We choose the two-site operator
\begin{equation}
  X_jX_{j+1}-Y_{j}Y_{j+1}=2(\sigma^+_j\sigma^+_{j+1}+\sigma^-_j\sigma^-_{j+1})
\end{equation}
which is sensitive to the creation/annihilation of pairs of particles.
For the mean value we find
\begin{equation}
\frac{ (\gamma \tanh(\alpha))^{L-1} -\gamma  \left(\gamma ^2-2\right) \tanh(\alpha) }{(\gamma \tanh(\alpha))^L +1}.
\label{XXYYmean}
\end{equation}
The infinite volume limit 
becomes
\begin{equation}
\vev{X_jX_{j+1}-Y_{j}Y_{j+1}}_{L\to\infty}=\gamma(2-\gamma^2)\tanh(\alpha).
\end{equation}
Having a non-zero mean value is a clear sign of the breaking of the
original $U(1)$-symmetry. 

\secc{Dynamics and Gibbs ensemble} We consider real time evolution from selected initial states, focusing on
\begin{equation}
  \label{rho0}
\rho(t=0)=  \rho_0(\beta)\equiv \frac{e^{\beta Q_0}}{(2\cosh \beta)^L}.
\end{equation}
These are steady states of the undeformed model ($\gamma=0$). They are product operators in real space and in the limit 
$\beta\to\pm\infty$ they
also include pure states obtained from the reference states with all spins up/down.

\begin{figure}[t]
  \centering
  \includegraphics{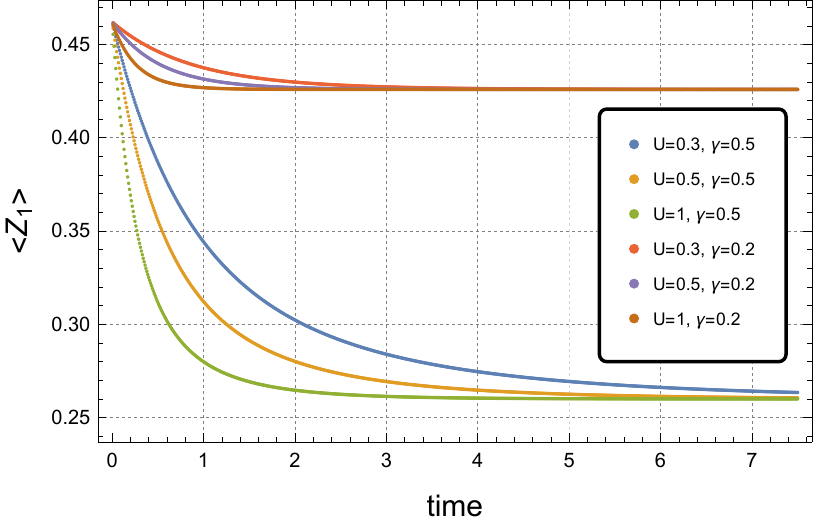}
  \caption{Time evolution of $\vev{Z_1(t)}$ from a selected initial density matrix $\rho_0(\beta)$ \eqref{rho0} with 
    $\beta=0.5$, in a finite volume $L=7$. We choose two 
    different deformation parameters $\gamma$ and three coupling strengths $U$. It is seen that the asymptotic values
    depend only on $\gamma$ and not on $U$, which influences only the speed of convergence. The asymptotic values agree
    with those predicted by the exact formula \eqref{Q0exact}, therefore they also confirm our postulate about the
    emergence of the Gibbs ensemble.}
  \label{fig:time}
\end{figure}

Due to the conserved charge $Q_\gamma$ we expect that the 
system has memory: the long time limit of the observables will depend on the initial state. On the other hand, since the emerging NESS are strong symmetries of the model, they are independent from the 
coupling $U$, therefore we expect that $U$ influences only the speed
of convergence towards them. This is confirmed by numerical computation of the real time dynamics for small volumes,
results are presented in Fig. \ref{fig:time}.

It is important to clarify the nature of the emerging steady states. 
Our Lindblad system has a single extensive conserved charge $Q_\gamma$. In analogy with
thermalization in an isolated system, we postulate that in large volumes the emerging steady states can be described by
a Gibbs Ensemble of the form 
\begin{equation}
  \rho_{\rm G}\sim e^{-\lambda Q_\gamma},
\end{equation}
such that $\lambda$ should be determined by the initial mean value:
\begin{equation}
  \text{Tr}\big(\rho_{\rm G} Q_\gamma\big)=\text{Tr} \big(\rho_0 Q_\gamma\big) .
\end{equation}
For the initial density matrices \eqref{rho0} this computation can be performed easily in the infinite volume limit,
yielding $\tanh(\lambda)=-(1-\gamma^2)\tanh(\beta)$. This result can be used to compute mean values of local observables
in the Gibbs Ensemble. We obtain for example the prediction
\begin{equation}
  \label{gibbsres}
\lim_{t\to\infty}  \vev{Z_j(t)}=\text{Tr}\big(\rho_{\rm G} Z_j\big)=(1-\gamma^2)^2 \tanh(\beta) .
\end{equation}

Remarkably, we also performed an exact finite volume computation to find the asymptotic mean values \cite{Note1}. For the
observable $Z_j$ we find
\begin{align}
\lim_{t\to\infty}\langle Z_j \rangle &= \frac{(\gamma ^2-1)^2  \tanh \beta  (1-2 \gamma ^L \tanh ^{L-2}\beta +\gamma ^{2L})}{(1-\gamma ^{2 L})^2} .\,
\label{Q0exact} 
\end{align}
These values are confirmed by the numerics at finite $L$. Furthermore, it is easy to take the large volume limit, and
for $0<\gamma<1$ we always recover \eqref{gibbsres}, thus confirming also our postulate about the Gibbs Ensemble.

\secc{Conclusions} We demonstrated that a Lindblad system with local jump operators can have quasi-local symmetries,
crucially affecting the real time dynamics. The steady states of our model were obtained from those of the ``Hubbard
Lindbladian'' after a similarity transformation with an MPO. Surprisingly, this similarity transformation is compatible
with local jump operators. Curiously we did not use the integrability of the Lindbladian, but the superintegrability of
the Hamiltonian did play  a crucial role. Perhaps integrability plays a hidden role in the derivation of
\eqref{Q0exact}.

Additional physical properties of the model, such as the Lindbladian gap could be computed from a full Bethe Ansatz
solution, which is not yet available.
Also, it would be interesting to consider analogous models with discrete time evolution
\cite{lindblad-circuit,lindblad-circuits-2}. This would open up the way towards the experimental realization of our findings.

\secc{Acknowledgments} We are thankful for useful discussions with Marko Ljubotina, Sanjay Moudgalya, Toma\v{z} Prosen,
Maksym Serbyn, Tibor Rakovszky. 
MdL was
supported by SFI, the Royal Society and the EPSRC for funding under
grants UF160578, RGF$\backslash$R1$\backslash$181011,
RGF$\backslash$EA$\backslash$180167 and 18/EPSRC/3590. CP was supported by RGF$\backslash$EA$\backslash$180167.


\begin{thebibliography}{39}%
\makeatletter
\providecommand \@ifxundefined [1]{%
 \@ifx{#1\undefined}
}%
\providecommand \@ifnum [1]{%
 \ifnum #1\expandafter \@firstoftwo
 \else \expandafter \@secondoftwo
 \fi
}%
\providecommand \@ifx [1]{%
 \ifx #1\expandafter \@firstoftwo
 \else \expandafter \@secondoftwo
 \fi
}%
\providecommand \natexlab [1]{#1}%
\providecommand \enquote  [1]{``#1''}%
\providecommand \bibnamefont  [1]{#1}%
\providecommand \bibfnamefont [1]{#1}%
\providecommand \citenamefont [1]{#1}%
\providecommand \href@noop [0]{\@secondoftwo}%
\providecommand \href [0]{\begingroup \@sanitize@url \@href}%
\providecommand \@href[1]{\@@startlink{#1}\@@href}%
\providecommand \@@href[1]{\endgroup#1\@@endlink}%
\providecommand \@sanitize@url [0]{\catcode `\\12\catcode `\$12\catcode
  `\&12\catcode `\#12\catcode `\^12\catcode `\_12\catcode `\%12\relax}%
\providecommand \@@startlink[1]{}%
\providecommand \@@endlink[0]{}%
\providecommand \url  [0]{\begingroup\@sanitize@url \@url }%
\providecommand \@url [1]{\endgroup\@href {#1}{\urlprefix }}%
\providecommand \urlprefix  [0]{URL }%
\providecommand \Eprint [0]{\href }%
\providecommand \doibase [0]{https://doi.org/}%
\providecommand \selectlanguage [0]{\@gobble}%
\providecommand \bibinfo  [0]{\@secondoftwo}%
\providecommand \bibfield  [0]{\@secondoftwo}%
\providecommand \translation [1]{[#1]}%
\providecommand \BibitemOpen [0]{}%
\providecommand \bibitemStop [0]{}%
\providecommand \bibitemNoStop [0]{.\EOS\space}%
\providecommand \EOS [0]{\spacefactor3000\relax}%
\providecommand \BibitemShut  [1]{\csname bibitem#1\endcsname}%
\let\auto@bib@innerbib\@empty
\bibitem [{\citenamefont {{De Palma}}\ \emph {et~al.}(2015)\citenamefont {{De
  Palma}}, \citenamefont {{Serafini}}, \citenamefont {{Giovannetti}},\ and\
  \citenamefont {{Cramer}}}]{necessity-of-eth}%
  \BibitemOpen
  \bibfield  {author} {\bibinfo {author} {\bibfnamefont {G.}~\bibnamefont {{De
  Palma}}}, \bibinfo {author} {\bibfnamefont {A.}~\bibnamefont {{Serafini}}},
  \bibinfo {author} {\bibfnamefont {V.}~\bibnamefont {{Giovannetti}}},\ and\
  \bibinfo {author} {\bibfnamefont {M.}~\bibnamefont {{Cramer}}},\ }\bibfield
  {title} {\bibinfo {title} {{Necessity of Eigenstate Thermalization}},\ }\href
  {https://doi.org/10.1103/PhysRevLett.115.220401} {\bibfield  {journal}
  {\bibinfo  {journal} {Phys. Rev. Lett.}\ }\textbf {\bibinfo {volume} {115}},\
  \bibinfo {eid} {220401} (\bibinfo {year} {2015})},\ \Eprint
  {https://arxiv.org/abs/1506.07265} {arXiv:1506.07265 [quant-ph]} \BibitemShut
  {NoStop}%
\bibitem [{\citenamefont {{Lange}}\ \emph {et~al.}(2017)\citenamefont
  {{Lange}}, \citenamefont {{Lenar{\v{c}}i{\v{c}}}},\ and\ \citenamefont
  {{Rosch}}}]{pumping}%
  \BibitemOpen
  \bibfield  {author} {\bibinfo {author} {\bibfnamefont {F.}~\bibnamefont
  {{Lange}}}, \bibinfo {author} {\bibfnamefont {Z.}~\bibnamefont
  {{Lenar{\v{c}}i{\v{c}}}}},\ and\ \bibinfo {author} {\bibfnamefont
  {A.}~\bibnamefont {{Rosch}}},\ }\bibfield  {title} {\bibinfo {title}
  {{Pumping approximately integrable systems}},\ }\href
  {https://doi.org/10.1038/ncomms15767} {\bibfield  {journal} {\bibinfo
  {journal} {Nature Communications}\ }\textbf {\bibinfo {volume} {8}},\
  \bibinfo {eid} {15767} (\bibinfo {year} {2017})},\ \Eprint
  {https://arxiv.org/abs/1608.03563} {arXiv:1608.03563 [cond-mat.str-el]}
  \BibitemShut {NoStop}%
\bibitem [{\citenamefont {Ikeda}\ and\ \citenamefont
  {Sato}(2020)}]{ness-cite1}%
  \BibitemOpen
  \bibfield  {author} {\bibinfo {author} {\bibfnamefont {T.~N.}\ \bibnamefont
  {Ikeda}}\ and\ \bibinfo {author} {\bibfnamefont {M.}~\bibnamefont {Sato}},\
  }\bibfield  {title} {\bibinfo {title} {General description for nonequilibrium
  steady states in periodically driven dissipative quantum systems},\
  }\bibfield  {journal} {\bibinfo  {journal} {Science Adv.}\ }\textbf {\bibinfo
  {volume} {6}},\ \href {https://doi.org/10.1126/sciadv.abb4019}
  {10.1126/sciadv.abb4019} (\bibinfo {year} {2020}),\ \Eprint
  {https://arxiv.org/abs/2003.02876} {arXiv:2003.02876 [cond-mat.stat-mech]}
  \BibitemShut {NoStop}%
\bibitem [{\citenamefont {Manzano}(2020)}]{lindblad-intro}%
  \BibitemOpen
  \bibfield  {author} {\bibinfo {author} {\bibfnamefont {D.}~\bibnamefont
  {Manzano}},\ }\bibfield  {title} {\bibinfo {title} {A short introduction to
  the lindblad master equation},\ }\href {https://doi.org/10.1063/1.5115323}
  {\bibfield  {journal} {\bibinfo  {journal} {AIP Advances}\ }\textbf {\bibinfo
  {volume} {10}},\ \bibinfo {pages} {025106} (\bibinfo {year} {2020})},\
  \Eprint {https://arxiv.org/abs/1906.04478} {arXiv:1906.04478 [quant-ph]}
  \BibitemShut {NoStop}%
\bibitem [{\citenamefont {{Nigro}}(2019)}]{lindblad-unique-1}%
  \BibitemOpen
  \bibfield  {author} {\bibinfo {author} {\bibfnamefont {D.}~\bibnamefont
  {{Nigro}}},\ }\bibfield  {title} {\bibinfo {title} {{On the uniqueness of the
  steady-state solution of the Lindblad-Gorini-Kossakowski-Sudarshan
  equation}},\ }\href {https://doi.org/10.1088/1742-5468/ab0c1c} {\bibfield
  {journal} {\bibinfo  {journal} {JSTAT}\ }\textbf {\bibinfo {volume}
  {4}}\bibfield  {number} {\bibinfo  {number} { (4)},\ \bibinfo {pages}
  {043202}},\ }\Eprint {https://arxiv.org/abs/1803.06279} {arXiv:1803.06279
  [quant-ph]} \BibitemShut {NoStop}%
\bibitem [{\citenamefont {{Albert}}(2018)}]{multiness-phd-thesis}%
  \BibitemOpen
  \bibfield  {author} {\bibinfo {author} {\bibfnamefont {V.~V.}\ \bibnamefont
  {{Albert}}},\ }\bibfield  {title} {\bibinfo {title} {{Lindbladians with
  multiple steady states: theory and applications}},\ }\href@noop {} {\bibfield
   {journal} {\bibinfo  {journal} {arXiv e-prints}\ } (\bibinfo {year}
  {2018})},\ \Eprint {https://arxiv.org/abs/1802.00010} {arXiv:1802.00010
  [quant-ph]} \BibitemShut {NoStop}%
\bibitem [{\citenamefont {{Nandkishore}}\ and\ \citenamefont
  {{Huse}}(2015)}]{huse-review}%
  \BibitemOpen
  \bibfield  {author} {\bibinfo {author} {\bibfnamefont {R.}~\bibnamefont
  {{Nandkishore}}}\ and\ \bibinfo {author} {\bibfnamefont {D.~A.}\ \bibnamefont
  {{Huse}}},\ }\bibfield  {title} {\bibinfo {title} {{Many body localization
  and thermalization in quantum statistical mechanics}},\ }\href
  {https://doi.org/10.1146/annurev-conmatphys-031214-014726} {\bibfield
  {journal} {\bibinfo  {journal} {Ann. Rev. Cond. Mat. Phys.}\ }\textbf
  {\bibinfo {volume} {6}},\ \bibinfo {pages} {15} (\bibinfo {year} {2015})},\
  \Eprint {https://arxiv.org/abs/1404.0686} {arXiv:1404.0686
  [cond-mat.stat-mech]} \BibitemShut {NoStop}%
\bibitem [{\citenamefont {{Essler}}\ and\ \citenamefont
  {{Fagotti}}(2016)}]{essler-fagotti-quench-review}%
  \BibitemOpen
  \bibfield  {author} {\bibinfo {author} {\bibfnamefont {F.~H.~L.}\
  \bibnamefont {{Essler}}}\ and\ \bibinfo {author} {\bibfnamefont
  {M.}~\bibnamefont {{Fagotti}}},\ }\bibfield  {title} {\bibinfo {title}
  {{Quench dynamics and relaxation in isolated integrable quantum spin
  chains}},\ }\href {https://doi.org/10.1088/1742-5468/2016/06/064002}
  {\bibfield  {journal} {\bibinfo  {journal} {J. Stat. Mech.}\ }\textbf
  {\bibinfo {volume} {6}},\ \bibinfo {pages} {064002} (\bibinfo {year}
  {2016})},\ \Eprint {https://arxiv.org/abs/1603.06452} {arXiv:1603.06452
  [cond-mat.quant-gas]} \BibitemShut {NoStop}%
\bibitem [{\citenamefont {{Moudgalya}}\ \emph {et~al.}(2022)\citenamefont
  {{Moudgalya}}, \citenamefont {{Bernevig}},\ and\ \citenamefont
  {{Regnault}}}]{hfrag-review}%
  \BibitemOpen
  \bibfield  {author} {\bibinfo {author} {\bibfnamefont {S.}~\bibnamefont
  {{Moudgalya}}}, \bibinfo {author} {\bibfnamefont {B.~A.}\ \bibnamefont
  {{Bernevig}}},\ and\ \bibinfo {author} {\bibfnamefont {N.}~\bibnamefont
  {{Regnault}}},\ }\bibfield  {title} {\bibinfo {title} {Quantum many-body
  scars and hilbert space fragmentation: a review of exact results},\ }\href
  {https://doi.org/10.1088/1361-6633/ac73a0} {\bibfield  {journal} {\bibinfo
  {journal} {Rep. Prog. Phys.}\ }\textbf {\bibinfo {volume} {85}},\ \bibinfo
  {pages} {086501} (\bibinfo {year} {2022})},\ \Eprint
  {https://arxiv.org/abs/2109.00548} {arXiv:2109.00548 [cond-mat.str-el]}
  \BibitemShut {NoStop}%
\bibitem [{\citenamefont {{Bu{\v{c}}a}}\ and\ \citenamefont
  {{Prosen}}(2012)}]{prosen-strong-symmetries}%
  \BibitemOpen
  \bibfield  {author} {\bibinfo {author} {\bibfnamefont {B.}~\bibnamefont
  {{Bu{\v{c}}a}}}\ and\ \bibinfo {author} {\bibfnamefont {T.}~\bibnamefont
  {{Prosen}}},\ }\bibfield  {title} {\bibinfo {title} {{A note on symmetry
  reductions of the Lindblad equation: transport in constrained open spin
  chains}},\ }\href {https://doi.org/10.1088/1367-2630/14/7/073007} {\bibfield
  {journal} {\bibinfo  {journal} {New J. Phys.}\ }\textbf {\bibinfo {volume}
  {14}},\ \bibinfo {eid} {073007} (\bibinfo {year} {2012})},\ \Eprint
  {https://arxiv.org/abs/1203.0943} {arXiv:1203.0943 [quant-ph]} \BibitemShut
  {NoStop}%
\bibitem [{\citenamefont {{Albert}}\ and\ \citenamefont
  {{Jiang}}(2014)}]{lindblad-symmetries}%
  \BibitemOpen
  \bibfield  {author} {\bibinfo {author} {\bibfnamefont {V.~V.}\ \bibnamefont
  {{Albert}}}\ and\ \bibinfo {author} {\bibfnamefont {L.}~\bibnamefont
  {{Jiang}}},\ }\bibfield  {title} {\bibinfo {title} {{Symmetries and conserved
  quantities in Lindblad master equations}},\ }\href
  {https://doi.org/10.1103/PhysRevA.89.022118} {\bibfield  {journal} {\bibinfo
  {journal} {Phys. Rev. A}\ }\textbf {\bibinfo {volume} {89}},\ \bibinfo {eid}
  {022118} (\bibinfo {year} {2014})},\ \Eprint
  {https://arxiv.org/abs/1310.1523} {arXiv:1310.1523 [quant-ph]} \BibitemShut
  {NoStop}%
\bibitem [{\citenamefont {{Ilievski}}\ and\ \citenamefont
  {{Prosen}}(2014)}]{enej-prosen-spin1degness}%
  \BibitemOpen
  \bibfield  {author} {\bibinfo {author} {\bibfnamefont {E.}~\bibnamefont
  {{Ilievski}}}\ and\ \bibinfo {author} {\bibfnamefont {T.}~\bibnamefont
  {{Prosen}}},\ }\bibfield  {title} {\bibinfo {title} {{Exact steady state
  manifold of a boundary driven spin-1 Lai-Sutherland chain}},\ }\href
  {https://doi.org/10.1016/j.nuclphysb.2014.03.016} {\bibfield  {journal}
  {\bibinfo  {journal} {Nucl. Phys. B}\ }\textbf {\bibinfo {volume} {882}},\
  \bibinfo {pages} {485} (\bibinfo {year} {2014})},\ \Eprint
  {https://arxiv.org/abs/1402.0342} {arXiv:1402.0342 [quant-ph]} \BibitemShut
  {NoStop}%
\bibitem [{\citenamefont {{Zhang}}\ \emph {et~al.}(2020)\citenamefont
  {{Zhang}}, \citenamefont {{Tindall}}, \citenamefont {{Mur-Petit}},
  \citenamefont {{Jaksch}},\ and\ \citenamefont
  {{Bu{\v{c}}a}}}]{buca-strong-symmetries}%
  \BibitemOpen
  \bibfield  {author} {\bibinfo {author} {\bibfnamefont {Z.}~\bibnamefont
  {{Zhang}}}, \bibinfo {author} {\bibfnamefont {J.}~\bibnamefont {{Tindall}}},
  \bibinfo {author} {\bibfnamefont {J.}~\bibnamefont {{Mur-Petit}}}, \bibinfo
  {author} {\bibfnamefont {D.}~\bibnamefont {{Jaksch}}},\ and\ \bibinfo
  {author} {\bibfnamefont {B.}~\bibnamefont {{Bu{\v{c}}a}}},\ }\bibfield
  {title} {\bibinfo {title} {{Stationary state degeneracy of open quantum
  systems with non-abelian symmetries}},\ }\href
  {https://doi.org/10.1088/1751-8121/ab88e3} {\bibfield  {journal} {\bibinfo
  {journal} {J. Phys. A}\ }\textbf {\bibinfo {volume} {53}},\ \bibinfo {eid}
  {215304} (\bibinfo {year} {2020})},\ \Eprint
  {https://arxiv.org/abs/1912.12185} {arXiv:1912.12185 [quant-ph]} \BibitemShut
  {NoStop}%
\bibitem [{\citenamefont {{Ilievski}}\ \emph {et~al.}(2016)\citenamefont
  {{Ilievski}}, \citenamefont {{Medenjak}}, \citenamefont {{Prosen}},\ and\
  \citenamefont {{Zadnik}}}]{prosen-enej-quasi-local-review}%
  \BibitemOpen
  \bibfield  {author} {\bibinfo {author} {\bibfnamefont {E.}~\bibnamefont
  {{Ilievski}}}, \bibinfo {author} {\bibfnamefont {M.}~\bibnamefont
  {{Medenjak}}}, \bibinfo {author} {\bibfnamefont {T.}~\bibnamefont
  {{Prosen}}},\ and\ \bibinfo {author} {\bibfnamefont {L.}~\bibnamefont
  {{Zadnik}}},\ }\bibfield  {title} {\bibinfo {title} {{Quasilocal charges in
  integrable lattice systems}},\ }\href
  {https://doi.org/10.1088/1742-5468/2016/06/064008} {\bibfield  {journal}
  {\bibinfo  {journal} {J. Stat. Mech.}\ }\textbf {\bibinfo {volume} {6}},\
  \bibinfo {pages} {064008} (\bibinfo {year} {2016})},\ \Eprint
  {https://arxiv.org/abs/1603.00440} {arXiv:1603.00440 [cond-mat.stat-mech]}
  \BibitemShut {NoStop}%
\bibitem [{\citenamefont {{Ilievski}}\ \emph {et~al.}(2015)\citenamefont
  {{Ilievski}}, \citenamefont {{De Nardis}}, \citenamefont {{Wouters}},
  \citenamefont {{Caux}}, \citenamefont {{Essler}},\ and\ \citenamefont
  {{Prosen}}}]{JS-CGGE}%
  \BibitemOpen
  \bibfield  {author} {\bibinfo {author} {\bibfnamefont {E.}~\bibnamefont
  {{Ilievski}}}, \bibinfo {author} {\bibfnamefont {J.}~\bibnamefont {{De
  Nardis}}}, \bibinfo {author} {\bibfnamefont {B.}~\bibnamefont {{Wouters}}},
  \bibinfo {author} {\bibfnamefont {J.-S.}\ \bibnamefont {{Caux}}}, \bibinfo
  {author} {\bibfnamefont {F.~H.~L.}\ \bibnamefont {{Essler}}},\ and\ \bibinfo
  {author} {\bibfnamefont {T.}~\bibnamefont {{Prosen}}},\ }\bibfield  {title}
  {\bibinfo {title} {{Complete Generalized Gibbs Ensembles in an Interacting
  Theory}},\ }\href {https://doi.org/10.1103/PhysRevLett.115.157201} {\bibfield
   {journal} {\bibinfo  {journal} {Phys. Rev. Lett.}\ }\textbf {\bibinfo
  {volume} {115}},\ \bibinfo {eid} {157201} (\bibinfo {year} {2015})},\ \Eprint
  {https://arxiv.org/abs/1507.02993} {arXiv:1507.02993 [quant-ph]} \BibitemShut
  {NoStop}%
\bibitem [{\citenamefont {Papi{\'{c}}}(2022)}]{fragmentation-scars-review-2}%
  \BibitemOpen
  \bibfield  {author} {\bibinfo {author} {\bibfnamefont {Z.}~\bibnamefont
  {Papi{\'{c}}}},\ }\bibfield  {title} {\bibinfo {title} {Weak ergodicity
  breaking through the lens of quantum entanglement},\ }in\ \href
  {https://doi.org/10.1007/978-3-031-03998-0_13} {\emph {\bibinfo {booktitle}
  {Entanglement in Spin Chains: From Theory to Quantum Technology
  Applications}}},\ \bibinfo {editor} {edited by\ \bibinfo {editor}
  {\bibfnamefont {A.}~\bibnamefont {Bayat}}, \bibinfo {editor} {\bibfnamefont
  {S.}~\bibnamefont {Bose}},\ and\ \bibinfo {editor} {\bibfnamefont
  {H.}~\bibnamefont {Johannesson}}}\ (\bibinfo  {publisher} {Springer
  International Publishing},\ \bibinfo {address} {Cham},\ \bibinfo {year}
  {2022})\ pp.\ \bibinfo {pages} {341--395},\ \Eprint
  {https://arxiv.org/abs/2108.03460} {arXiv:2108.03460 [cond-mat.quant-gas]}
  \BibitemShut {NoStop}%
\bibitem [{\citenamefont {Lindblad}(1976)}]{lindblad-eredeti}%
  \BibitemOpen
  \bibfield  {author} {\bibinfo {author} {\bibfnamefont {G.}~\bibnamefont
  {Lindblad}},\ }\bibfield  {title} {\bibinfo {title} {On the generators of
  quantum dynamical semigroups},\ }\href {https://doi.org/10.1007/BF01608499}
  {\bibfield  {journal} {\bibinfo  {journal} {Commun.Math. Phys.}\ }\textbf
  {\bibinfo {volume} {48}},\ \bibinfo {pages} {119} (\bibinfo {year}
  {1976})}\BibitemShut {NoStop}%
\bibitem [{\citenamefont {Breuer}\ \emph {et~al.}(2002)\citenamefont {Breuer},
  \citenamefont {Petruccione} \emph {et~al.}}]{breuer2002theory}%
  \BibitemOpen
  \bibfield  {author} {\bibinfo {author} {\bibfnamefont {H.-P.}\ \bibnamefont
  {Breuer}}, \bibinfo {author} {\bibfnamefont {F.}~\bibnamefont {Petruccione}},
  \emph {et~al.},\ }\href@noop {} {\emph {\bibinfo {title} {The theory of open
  quantum systems}}}\ (\bibinfo  {publisher} {Oxford University Press on
  Demand},\ \bibinfo {year} {2002})\BibitemShut {NoStop}%
\bibitem [{\citenamefont {{Medvedyeva}}\ \emph {et~al.}(2016)\citenamefont
  {{Medvedyeva}}, \citenamefont {{Essler}},\ and\ \citenamefont
  {{Prosen}}}]{essler-prosen-lindblad}%
  \BibitemOpen
  \bibfield  {author} {\bibinfo {author} {\bibfnamefont {M.~V.}\ \bibnamefont
  {{Medvedyeva}}}, \bibinfo {author} {\bibfnamefont {F.~H.~L.}\ \bibnamefont
  {{Essler}}},\ and\ \bibinfo {author} {\bibfnamefont {T.}~\bibnamefont
  {{Prosen}}},\ }\bibfield  {title} {\bibinfo {title} {{Exact Bethe Ansatz
  Spectrum of a Tight-Binding Chain with Dephasing Noise}},\ }\href
  {https://doi.org/10.1103/PhysRevLett.117.137202} {\bibfield  {journal}
  {\bibinfo  {journal} {Phys. Rev. Lett.}\ }\textbf {\bibinfo {volume} {117}},\
  \bibinfo {eid} {137202} (\bibinfo {year} {2016})},\ \Eprint
  {https://arxiv.org/abs/1606.09122} {arXiv:1606.09122 [quant-ph]} \BibitemShut
  {NoStop}%
\bibitem [{\citenamefont {Vernier}\ \emph {et~al.}(2019)\citenamefont
  {Vernier}, \citenamefont {O’Brien},\ and\ \citenamefont
  {Fendley}}]{ericZn}%
  \BibitemOpen
  \bibfield  {author} {\bibinfo {author} {\bibfnamefont {E.}~\bibnamefont
  {Vernier}}, \bibinfo {author} {\bibfnamefont {E.}~\bibnamefont {O’Brien}},\
  and\ \bibinfo {author} {\bibfnamefont {P.}~\bibnamefont {Fendley}},\
  }\bibfield  {title} {\bibinfo {title} {Onsager symmetries in $u(1)$-invariant
  clock models},\ }\href {https://doi.org/10.1088/1742-5468/ab11c0} {\bibfield
  {journal} {\bibinfo  {journal} {J. Stat. Mech.}\ }\textbf {\bibinfo {volume}
  {2019}},\ \bibinfo {pages} {043107} (\bibinfo {year} {2019})},\ \Eprint
  {https://arxiv.org/abs/1812.09091} {arXiv:1812.09091} \BibitemShut {NoStop}%
\bibitem [{\citenamefont {Lieb}\ \emph {et~al.}(1961)\citenamefont {Lieb},
  \citenamefont {Schultz},\ and\ \citenamefont {Mattis}}]{XX-original}%
  \BibitemOpen
  \bibfield  {author} {\bibinfo {author} {\bibfnamefont {E.}~\bibnamefont
  {Lieb}}, \bibinfo {author} {\bibfnamefont {T.}~\bibnamefont {Schultz}},\ and\
  \bibinfo {author} {\bibfnamefont {D.}~\bibnamefont {Mattis}},\ }\bibfield
  {title} {\bibinfo {title} {Two soluble models of an antiferromagnetic
  chain},\ }\href {https://doi.org/10.1016/0003-4916(61)90115-4} {\bibfield
  {journal} {\bibinfo  {journal} {Ann. Phys.}\ }\textbf {\bibinfo {volume}
  {16}},\ \bibinfo {pages} {407} (\bibinfo {year} {1961})}\BibitemShut
  {NoStop}%
\bibitem [{\citenamefont {{de Leeuw}}\ \emph {et~al.}(2021)\citenamefont {{de
  Leeuw}}, \citenamefont {{Paletta}},\ and\ \citenamefont
  {{Pozsgay}}}]{sajat-lindblad}%
  \BibitemOpen
  \bibfield  {author} {\bibinfo {author} {\bibfnamefont {M.}~\bibnamefont {{de
  Leeuw}}}, \bibinfo {author} {\bibfnamefont {C.}~\bibnamefont {{Paletta}}},\
  and\ \bibinfo {author} {\bibfnamefont {B.}~\bibnamefont {{Pozsgay}}},\
  }\bibfield  {title} {\bibinfo {title} {{Constructing Integrable Lindblad
  Superoperators}},\ }\href {https://doi.org/10.1103/PhysRevLett.126.240403}
  {\bibfield  {journal} {\bibinfo  {journal} {Phys. Rev. Lett.}\ }\textbf
  {\bibinfo {volume} {126}},\ \bibinfo {eid} {240403} (\bibinfo {year}
  {2021})},\ \Eprint {https://arxiv.org/abs/2101.08279} {arXiv:2101.08279
  [cond-mat.stat-mech]} \BibitemShut {NoStop}%
\bibitem [{\citenamefont {{Ziolkowska}}\ and\ \citenamefont
  {{Essler}}(2019)}]{essler-lindblad-review}%
  \BibitemOpen
  \bibfield  {author} {\bibinfo {author} {\bibfnamefont {A.~A.}\ \bibnamefont
  {{Ziolkowska}}}\ and\ \bibinfo {author} {\bibfnamefont {F.~H.~L.}\
  \bibnamefont {{Essler}}},\ }\bibfield  {title} {\bibinfo {title}
  {{Yang-Baxter integrable Lindblad equations}},\ }\href
  {https://doi.org/10.21468/SciPostPhys.8.3.044} {\bibfield  {journal}
  {\bibinfo  {journal} {SciPost Phys.}\ }\textbf {\bibinfo {volume} {8}},\
  \bibinfo {pages} {044} (\bibinfo {year} {2019})},\ \Eprint
  {https://arxiv.org/abs/1911.04883} {arXiv:1911.04883 [cond-mat.stat-mech]}
  \BibitemShut {NoStop}%
\bibitem [{\citenamefont {{de Leeuw}}\ \emph {et~al.}(2023)\citenamefont {{de
  Leeuw}}, \citenamefont {{Paletta}},\ and\ \citenamefont
  {{Pozsgay}}}]{sajat-hubbard}%
  \BibitemOpen
  \bibfield  {author} {\bibinfo {author} {\bibfnamefont {M.}~\bibnamefont {{de
  Leeuw}}}, \bibinfo {author} {\bibfnamefont {C.}~\bibnamefont {{Paletta}}},\
  and\ \bibinfo {author} {\bibfnamefont {B.}~\bibnamefont {{Pozsgay}}},\
  }\bibfield  {title} {\bibinfo {title} {{A range three elliptic deformation of
  the Hubbard model}},\ }\href@noop {} {\bibfield  {journal} {\bibinfo
  {journal} {arXiv e-prints}\ } (\bibinfo {year} {2023})},\ \Eprint
  {https://arxiv.org/abs/2301.01612} {arXiv:2301.01612 [cond-mat.stat-mech]}
  \BibitemShut {NoStop}%
\bibitem [{\citenamefont {{Miller}}\ \emph {et~al.}(2013)\citenamefont
  {{Miller}}, \citenamefont {{Post}},\ and\ \citenamefont
  {{Winternitz}}}]{superintegrability}%
  \BibitemOpen
  \bibfield  {author} {\bibinfo {author} {\bibfnamefont {J.}~\bibnamefont
  {{Miller}}, \bibfnamefont {Willard}}, \bibinfo {author} {\bibfnamefont
  {S.}~\bibnamefont {{Post}}},\ and\ \bibinfo {author} {\bibfnamefont
  {P.}~\bibnamefont {{Winternitz}}},\ }\bibfield  {title} {\bibinfo {title}
  {{Classical and quantum superintegrability with applications}},\ }\href
  {https://doi.org/10.1088/1751-8113/46/42/423001} {\bibfield  {journal}
  {\bibinfo  {journal} {J. Phys. A}\ }\textbf {\bibinfo {volume} {46}},\
  \bibinfo {eid} {423001} (\bibinfo {year} {2013})},\ \Eprint
  {https://arxiv.org/abs/1309.2694} {arXiv:1309.2694 [math-ph]} \BibitemShut
  {NoStop}%
\bibitem [{Note1()}]{Note1}%
  \BibitemOpen
  \bibinfo {note} {Supplemental Materials to ``Hidden quasi-local charges and
  Gibbs ensemble in a Lindblad system''}\BibitemShut {NoStop}%
\bibitem [{\citenamefont
  {{Prosen}}(2011{\natexlab{a}})}]{prosen-boundary-lindblad-1}%
  \BibitemOpen
  \bibfield  {author} {\bibinfo {author} {\bibfnamefont {T.}~\bibnamefont
  {{Prosen}}},\ }\bibfield  {title} {\bibinfo {title} {{Open XXZ Spin Chain:
  Nonequilibrium Steady State and a Strict Bound on Ballistic Transport}},\
  }\href {https://doi.org/10.1103/PhysRevLett.106.217206} {\bibfield  {journal}
  {\bibinfo  {journal} {Phys. Rev. Lett.}\ }\textbf {\bibinfo {volume} {106}},\
  \bibinfo {eid} {217206} (\bibinfo {year} {2011}{\natexlab{a}})},\ \Eprint
  {https://arxiv.org/abs/1103.1350} {arXiv:1103.1350 [cond-mat.str-el]}
  \BibitemShut {NoStop}%
\bibitem [{\citenamefont {{Prosen}}\ \emph {et~al.}(2013)\citenamefont
  {{Prosen}}, \citenamefont {{Ilievski}},\ and\ \citenamefont
  {{Popkov}}}]{prosen-exterior-lindblad}%
  \BibitemOpen
  \bibfield  {author} {\bibinfo {author} {\bibfnamefont {T.}~\bibnamefont
  {{Prosen}}}, \bibinfo {author} {\bibfnamefont {E.}~\bibnamefont
  {{Ilievski}}},\ and\ \bibinfo {author} {\bibfnamefont {V.}~\bibnamefont
  {{Popkov}}},\ }\bibfield  {title} {\bibinfo {title} {{Exterior integrability:
  Yang-Baxter form of non-equilibrium steady-state density operator}},\ }\href
  {https://doi.org/10.1088/1367-2630/15/7/073051} {\bibfield  {journal}
  {\bibinfo  {journal} {New J. Physi.}\ }\textbf {\bibinfo {volume} {15}},\
  \bibinfo {eid} {073051} (\bibinfo {year} {2013})},\ \Eprint
  {https://arxiv.org/abs/1304.7944} {arXiv:1304.7944 [math-ph]} \BibitemShut
  {NoStop}%
\bibitem [{\citenamefont
  {{Prosen}}(2011{\natexlab{b}})}]{prosen-boundary-lindblad-2}%
  \BibitemOpen
  \bibfield  {author} {\bibinfo {author} {\bibfnamefont {T.}~\bibnamefont
  {{Prosen}}},\ }\bibfield  {title} {\bibinfo {title} {{Exact Nonequilibrium
  Steady State of a Strongly Driven Open XXZ Chain}},\ }\href
  {https://doi.org/10.1103/PhysRevLett.107.137201} {\bibfield  {journal}
  {\bibinfo  {journal} {Phys. Rev. Lett.}\ }\textbf {\bibinfo {volume} {107}},\
  \bibinfo {eid} {137201} (\bibinfo {year} {2011}{\natexlab{b}})},\ \Eprint
  {https://arxiv.org/abs/1106.2978} {arXiv:1106.2978 [quant-ph]} \BibitemShut
  {NoStop}%
\bibitem [{\citenamefont {{Jones}}\ \emph {et~al.}(2021)\citenamefont
  {{Jones}}, \citenamefont {{Bibo}}, \citenamefont {{Jobst}}, \citenamefont
  {{Pollmann}}, \citenamefont {{Smith}},\ and\ \citenamefont
  {{Verresen}}}]{MPS-cluster-model}%
  \BibitemOpen
  \bibfield  {author} {\bibinfo {author} {\bibfnamefont {N.~G.}\ \bibnamefont
  {{Jones}}}, \bibinfo {author} {\bibfnamefont {J.}~\bibnamefont {{Bibo}}},
  \bibinfo {author} {\bibfnamefont {B.}~\bibnamefont {{Jobst}}}, \bibinfo
  {author} {\bibfnamefont {F.}~\bibnamefont {{Pollmann}}}, \bibinfo {author}
  {\bibfnamefont {A.}~\bibnamefont {{Smith}}},\ and\ \bibinfo {author}
  {\bibfnamefont {R.}~\bibnamefont {{Verresen}}},\ }\bibfield  {title}
  {\bibinfo {title} {{Skeleton of matrix-product-state-solvable models
  connecting topological phases of matter}},\ }\href
  {https://doi.org/10.1103/PhysRevResearch.3.033265} {\bibfield  {journal}
  {\bibinfo  {journal} {Phys. Rev. Res.}\ }\textbf {\bibinfo {volume} {3}},\
  \bibinfo {eid} {033265} (\bibinfo {year} {2021})},\ \Eprint
  {https://arxiv.org/abs/2105.12143} {arXiv:2105.12143 [cond-mat.str-el]}
  \BibitemShut {NoStop}%
\bibitem [{\citenamefont {Peschel}\ and\ \citenamefont
  {Emery}(1981)}]{peschel-emery}%
  \BibitemOpen
  \bibfield  {author} {\bibinfo {author} {\bibfnamefont {I.}~\bibnamefont
  {Peschel}}\ and\ \bibinfo {author} {\bibfnamefont {V.}~\bibnamefont
  {Emery}},\ }\bibfield  {title} {\bibinfo {title} {Calculation of spin
  correlations in two-dimensional ising systems from one-dimensional kinetic
  models},\ }\href {https://doi.org/10.1007/BF01297524} {\bibfield  {journal}
  {\bibinfo  {journal} {Z. Physik B}\ }\textbf {\bibinfo {volume} {43}},\
  \bibinfo {pages} {241} (\bibinfo {year} {1981})}\BibitemShut {NoStop}%
\bibitem [{\citenamefont {{Wouters}}\ \emph {et~al.}(2021)\citenamefont
  {{Wouters}}, \citenamefont {{Katsura}},\ and\ \citenamefont
  {{Schuricht}}}]{frustration-witten}%
  \BibitemOpen
  \bibfield  {author} {\bibinfo {author} {\bibfnamefont {J.}~\bibnamefont
  {{Wouters}}}, \bibinfo {author} {\bibfnamefont {H.}~\bibnamefont
  {{Katsura}}},\ and\ \bibinfo {author} {\bibfnamefont {D.}~\bibnamefont
  {{Schuricht}}},\ }\bibfield  {title} {\bibinfo {title} {{Interrelations among
  frustration-free models via Witten's conjugation}},\ }\href
  {https://doi.org/10.21468/SciPostPhysCore.4.4.027} {\bibfield  {journal}
  {\bibinfo  {journal} {SciPost Phys. Core}\ }\textbf {\bibinfo {volume} {4}},\
  \bibinfo {pages} {027} (\bibinfo {year} {2021})},\ \Eprint
  {https://arxiv.org/abs/2005.12825} {arXiv:2005.12825 [cond-mat.str-el]}
  \BibitemShut {NoStop}%
\bibitem [{Note2()}]{Note2}%
  \BibitemOpen
  \bibinfo {note} {We would like to clarify that we are referring about the
  Hermiticity property of the action of the superoperator $\protect
  \mathaccentV {tilde}07E{\protect \mathcal {L}}$. This is not the same as
  Hermiticity or anti-Hermiticity of the commutator $[\rho ,H]$.}\BibitemShut
  {Stop}%
\bibitem [{Note3()}]{Note3}%
  \BibitemOpen
  \bibinfo {note} {Closely related notions of a Lindbladian scar appeared in
  \cite {buca-lindblad-scar,lindblad-scar-embed}, see also \cite
  {lindblad-scar-szeru}.}\BibitemShut {Stop}%
\bibitem [{\citenamefont {{S{\'a}}}\ \emph {et~al.}(2021)\citenamefont
  {{S{\'a}}}, \citenamefont {{Ribeiro}},\ and\ \citenamefont
  {{Prosen}}}]{lindblad-circuit}%
  \BibitemOpen
  \bibfield  {author} {\bibinfo {author} {\bibfnamefont {L.}~\bibnamefont
  {{S{\'a}}}}, \bibinfo {author} {\bibfnamefont {P.}~\bibnamefont
  {{Ribeiro}}},\ and\ \bibinfo {author} {\bibfnamefont {T.}~\bibnamefont
  {{Prosen}}},\ }\bibfield  {title} {\bibinfo {title} {{Integrable Non-unitary
  Open Quantum Circuits}},\ }\href
  {https://doi.org/10.1103/PhysRevB.103.115132} {\bibfield  {journal} {\bibinfo
   {journal} {Phys. Rev. B}\ }\textbf {\bibinfo {volume} {103}},\ \bibinfo
  {pages} {115132} (\bibinfo {year} {2021})},\ \Eprint
  {https://arxiv.org/abs/2011.06565} {arXiv:2011.06565 [cond-mat.stat-mech]}
  \BibitemShut {NoStop}%
\bibitem [{\citenamefont {{Su}}\ and\ \citenamefont
  {{Martin}}(2022)}]{lindblad-circuits-2}%
  \BibitemOpen
  \bibfield  {author} {\bibinfo {author} {\bibfnamefont {L.}~\bibnamefont
  {{Su}}}\ and\ \bibinfo {author} {\bibfnamefont {I.}~\bibnamefont
  {{Martin}}},\ }\bibfield  {title} {\bibinfo {title} {{Integrable nonunitary
  quantum circuits}},\ }\href {https://doi.org/10.1103/PhysRevB.106.134312}
  {\bibfield  {journal} {\bibinfo  {journal} {Phys. Rev. B}\ }\textbf {\bibinfo
  {volume} {106}},\ \bibinfo {eid} {134312} (\bibinfo {year} {2022})},\ \Eprint
  {https://arxiv.org/abs/2205.13483} {arXiv:2205.13483 [cond-mat.str-el]}
  \BibitemShut {NoStop}%
\bibitem [{\citenamefont {{Bu{\v{c}}a}}\ \emph {et~al.}(2019)\citenamefont
  {{Bu{\v{c}}a}}, \citenamefont {{Tindall}},\ and\ \citenamefont
  {{Jaksch}}}]{buca-lindblad-scar}%
  \BibitemOpen
  \bibfield  {author} {\bibinfo {author} {\bibfnamefont {B.}~\bibnamefont
  {{Bu{\v{c}}a}}}, \bibinfo {author} {\bibfnamefont {J.}~\bibnamefont
  {{Tindall}}},\ and\ \bibinfo {author} {\bibfnamefont {D.}~\bibnamefont
  {{Jaksch}}},\ }\bibfield  {title} {\bibinfo {title} {{Non-stationary coherent
  quantum many-body dynamics through dissipation}},\ }\href
  {https://doi.org/10.1038/s41467-019-09757-y} {\bibfield  {journal} {\bibinfo
  {journal} {Nat. Comm.}\ }\textbf {\bibinfo {volume} {10}},\ \bibinfo {eid}
  {1730} (\bibinfo {year} {2019})},\ \Eprint {https://arxiv.org/abs/1804.06744}
  {arXiv:1804.06744 [quant-ph]} \BibitemShut {NoStop}%
\bibitem [{\citenamefont {{Wang}}\ \emph {et~al.}(2023)\citenamefont {{Wang}},
  \citenamefont {{Yuan}}, \citenamefont {{Zhang}}, \citenamefont {{Wang}},
  \citenamefont {{Deng}},\ and\ \citenamefont {{Duan}}}]{lindblad-scar-embed}%
  \BibitemOpen
  \bibfield  {author} {\bibinfo {author} {\bibfnamefont {H.-R.}\ \bibnamefont
  {{Wang}}}, \bibinfo {author} {\bibfnamefont {D.}~\bibnamefont {{Yuan}}},
  \bibinfo {author} {\bibfnamefont {S.-Y.}\ \bibnamefont {{Zhang}}}, \bibinfo
  {author} {\bibfnamefont {Z.}~\bibnamefont {{Wang}}}, \bibinfo {author}
  {\bibfnamefont {D.-L.}\ \bibnamefont {{Deng}}},\ and\ \bibinfo {author}
  {\bibfnamefont {L.~M.}\ \bibnamefont {{Duan}}},\ }\bibfield  {title}
  {\bibinfo {title} {{Embedding Quantum Many-Body Scars into Decoherence-Free
  Subspaces}},\ }\href@noop {} {\bibfield  {journal} {\bibinfo  {journal}
  {arXiv e-prints}\ } (\bibinfo {year} {2023})},\ \Eprint
  {https://arxiv.org/abs/2304.08515} {arXiv:2304.08515 [quant-ph]} \BibitemShut
  {NoStop}%
\bibitem [{\citenamefont {{Chen}}\ \emph {et~al.}(2022)\citenamefont {{Chen}},
  \citenamefont {{Chen}},\ and\ \citenamefont {{Zhu}}}]{lindblad-scar-szeru}%
  \BibitemOpen
  \bibfield  {author} {\bibinfo {author} {\bibfnamefont {Q.}~\bibnamefont
  {{Chen}}}, \bibinfo {author} {\bibfnamefont {S.~A.}\ \bibnamefont {{Chen}}},\
  and\ \bibinfo {author} {\bibfnamefont {Z.}~\bibnamefont {{Zhu}}},\ }\bibfield
   {title} {\bibinfo {title} {{Weak Ergodicity Breaking in Non-Hermitian
  Many-body Systems}},\ }\bibfield  {journal} {\bibinfo  {journal} {arXiv
  e-prints}\ }\href {https://doi.org/10.48550/arXiv.2202.08638}
  {10.48550/arXiv.2202.08638} (\bibinfo {year} {2022}),\ \Eprint
  {https://arxiv.org/abs/2202.08638} {arXiv:2202.08638 [cond-mat.quant-gas]}
  \BibitemShut {NoStop}%
\end{thebibliography}

%

\end{document}


\begin{center}
  \textbf{\large Supplemental Material to ``Hidden quasi-local charges and Gibbs ensemble in a Lindblad system''}
  \medskip

\end{center}
\setcounter{equation}{0}
\setcounter{figure}{0}
\setcounter{table}{0}
\setcounter{page}{1}
\makeatletter
\renewcommand{\theequation}{S\arabic{equation}}
\renewcommand{\thefigure}{S\arabic{figure}}
\renewcommand{\bibnumfmt}[1]{[S#1]}
\renewcommand{\citenumfont}[1]{S#1}

\section{Construction of the NESS using the conserved charges of the Hamiltonian}

In this section we show how the deformed Linbladian considered in the main text as well as the corresponding $L+1$ NESS can be constructed as a deformation of the $\gamma=0$ case (imaginary coupling Hubbard model) using the conserved charges of the Hamiltonian 
(5) in main text. 

\subsection{Conserved charges of the Hamiltonian}

The Hamiltonian (5) commutes with an extensive set of charges which we label as $[ab]_m$, where $a$ and $b$ can take the label $X$ or $Y$, and for $m \geq 0$, 
\begin{equation} 
[ab]_m \equiv \sum_{j=1}^L a_j \left( \prod_{1\leq k<m} Z_{j+k}   \right) b_{j+m}  \,,
\label{conservedcharges} 
\end{equation} 
where $X_j$, $Y_j$ and $Z_j$ are the Pauli matrices acting on site $j$ of the spin chain.

In the fermionic formulation of the model, detailed in the main text, the corresponding charges are the set of all possible translationally invariant fermion bilinears.

An extensive set of local charges usually signals integrability, and indeed the Hamiltonian (5) can be related to the well-known integrable XX Hamiltonian (see eq. (2) in main text) by a  homogeneous twist along the chain. It is in fact {\it superintegrable}, as the charges $[ab]_m$ form various families which in turn do not commute with one another.
 For instance, the sets of charges $\{[XY]_m\}$ and $\{[YX]_n\}$ commute with one another, but only the combinations $\{[XY]_m - [YX]_m \}$ commute with the charges $\{[XX]_n\}$ or $\{[YY]_n\}$. We also introduce the charge 
 \begin{equation}
\mathcal{Z} =  \prod_{j=1}^L Z_{j}  \,,
 \end{equation}
which commutes with the Hamiltonian as well as with all the charges $[ab]_m$.

\subsection{The operator $T(\gamma)$}

As usual when dealing with quantum integrable models, families of mutually conserved charges can be generated by a matrix-product operator (MPO) called the transfer matrix.
Introduce the following MPO  
\begin{equation}
T(\gamma) = \mathrm{Tr}_{\mathcal{A}} ( A_{L}(\gamma) A_{L-1}(\gamma)  \ldots  A_1(\gamma) ) \,,
\end{equation}
where the ancillary space $\mathcal{A}$ has dimension 2, and where the matrices $A_j(\gamma)$ are defined as  
\begin{equation}
{A}_j(\gamma)
 = 
\left(  
\begin{array}{cc}
\frac{\sqrt{1-\gamma} + \sqrt{1+\gamma}Z_j}{2} & \frac{\sqrt{1+\gamma} X_j -  i \sqrt{1-\gamma}Y_j}{2} \\
\frac{\sqrt{1-\gamma} X_j +  i \sqrt{1+\gamma}Y_j}{2} &  \frac{ \sqrt{1+\gamma} - \sqrt{1-\gamma}Z_j}{2} \\ 
\end{array}
 \right)  \,.
\end{equation}
The matrices $T(\gamma)$ commute with one another for different $\gamma$, as can be traced back to known integrability properties of the XX chain (more precisely, they correspond to transfer matrices based on cyclic representations of the quantum group $U_q(sl_2)$ at $q=i$), and admit the following series expansion around $\gamma =0$ : 
\begin{equation}
T(\gamma) = \mathcal{U} \exp(\mathcal{G}(\gamma)) \,,
\label{expansionofT}
\end{equation} 
 where $\mathcal{U}$ is the one-site discrete translation operator, and 
\begin{equation}
\mathcal{G}(\gamma) = i \sum_{m \geq 1} \frac{\gamma^m}{2m} [YX]_m  \,. 
\label{calGdef}
\end{equation}
We emphasize that the expansion \eqref{calGdef} holds at all orders, even for a system for finite size $L$, as can be checked by computing explicitly the successive logarithmic derivatives of $T(\gamma)$ at $\gamma=0$. For $L\to \infty$, the series \eqref{calGdef} defines a quasi-local operator for $|\gamma|<1$. For finite $L$, it can be further rearranged using the properties : $[YX]_{m+L} = - \mathcal{Z} [YX]_m$ for $m\geq 1$, and $[YX]_L = -i L \mathcal{Z}$. A practical expression is 
\begin{equation}
\mathcal{G}(\gamma) = \frac{1}{2} \log(1+\gamma^L \mathcal{Z}) +  i \sum_{\substack{m \geq 1\\ m \notin L \mathbb{Z}} } \frac{\gamma^m}{2m} [YX]_m,    \,
\end{equation}
which splits between a first term which is hermitian, and an anti-hermitian part. From there, see in particular: 
\begin{equation}
\label{TTdag}
T(\gamma)T(\gamma)^{\dagger}  = 1 + \gamma^L \mathcal{Z} \,, 
\end{equation}
or, equivalently,
\begin{equation}
T(\gamma)^{-1} = \frac{1-\gamma^L \mathcal{Z}}{1-\gamma^{2L}} T(\gamma)^{\dagger}  \,.
\label{Tgammam1}
\end{equation}
The result of equation \eqref{TTdag} can be seen directly in the MPO formalism. We can write $T(\gamma)T(\gamma)^{\dagger}$ as a MPO of bond dimension 4, with ancillary space $\mathcal{A}\otimes \mathcal{A}$, namely 
\begin{equation}
T(\gamma)T(\gamma)^{\dagger} = \mathrm{Tr}_{\mathcal{A}\otimes \mathcal{A}} ( \mathcal{M}_{L}(\gamma) \mathcal{M}_{L-1}(\gamma)  \ldots  \mathcal{M}_1(\gamma) ) \,,
\label{TTdagMPO}
\end{equation}
where the $\mathcal{M}_j(\gamma)$ are $4\times 4$ matrices with entries expressed in terms of $X_j,Y_j,Z_j$.  
 The MPO is invariant under any change of basis performed in the ancillary space. Defining $V=e^{\frac{i \pi }{4} Y\otimes X}$, where $X$ and $Y$ are now Pauli matrices acting in each copy of the ancillary space $\mathcal{A}$, \eqref{TTdagMPO} can therefore be recovered by replacing the matrices $\mathcal{M}_j(\gamma)$ by $V \mathcal{M}_j(\gamma) V^{-1}$, which take the form 
 \begin{equation}
 \label{Mconjugated}
 V \mathcal{M}_j(\gamma) V^{-1} = 
 \left(
\begin{array}{cccc}
1 & \sqrt{1-\gamma^2} X_j & -i Y_j & - \sqrt{1-\gamma^2} Z_j  \\
0 & \gamma Z_j  & 0 & \gamma X_j \\
0 & 0 & 0 & 0 \\
0 & 0 & 0 & 0 
\end{array} 
  \right) \,.
 \end{equation}
From the block-diagonal form of \eqref{Mconjugated}, it is clear that after taking the trace in \eqref{TTdagMPO} only the two diagonal terms contribute, which give rise to the two terms in \eqref{TTdag}. 

\subsection{Construction of the NESS}

As we will now see, the transfer matrix $T(\gamma)$ can be used to construct the jump operators and NESS described in the main text. 
Let us study the transformation of the jump operators $\ell(j)$ (defined in eq. (6) of the main text) under conjugation by $T(\gamma)$.  
For this sake, it will be useful to introduce the following MPOs :

\begin{equation}
T(\gamma) B_j T(\gamma)^{\dagger} = \mathrm{Tr}_{\mathcal{A}\otimes \mathcal{A}} ( \mathcal{M}_{L}(\gamma) \ldots \mathcal{M}^B_{j}(\gamma)  \ldots  \mathcal{M}_1(\gamma) ) \,,
\label{TBTdagMPO}
\end{equation}
where $B\in\{X,Y,Z\}$. 
We find similarly: 
 \begin{equation}
 \label{MXconjugated}
 V \mathcal{M}^X_j(\gamma) V^{-1} = 
 \left(
\begin{array}{cccc}
0 & \gamma Z_j  & 0 & \gamma X_j \\
1 & \sqrt{1-\gamma^2} X_j & -i Y_j & - \sqrt{1-\gamma^2} Z_j  \\
0 & 0 & 0 & 0 \\
0 & 0 & 0 & 0 
\end{array} 
  \right) \,,
 \end{equation}

 \begin{equation}
 \label{MYconjugated}
 V \mathcal{M}^Y_j(\gamma) V^{-1} = 
 \left(
\begin{array}{cccc}
0 & 0 & 0 & 0 \\
0 & 0 & 0 & 0 \\
 i\sqrt{1-\gamma^2}  & i X_j &  \sqrt{1-\gamma^2}  Y_j & -i Z_j  \\
 \gamma Y_j  & 0  & -i \gamma  & 0 \\
\end{array} 
  \right) \,,
 \end{equation}
 
  \begin{equation}
 \label{MZconjugated}
 V \mathcal{M}^Z_j(\gamma) V^{-1} = 
 \left(
\begin{array}{cccc}
0 & 0 & 0 & 0 \\
0 & 0 & 0 & 0 \\
i \gamma Y_j  & 0  & \gamma  & 0 \\
 -\sqrt{1-\gamma^2}  & - X_j & i \sqrt{1-\gamma^2}  Y_j &  Z_j  \\
\end{array} 
  \right) \,.
 \end{equation}

 It will also be useful, for practical calculations, to introduce $\mathcal{M}^{(\alpha)}_j = \cosh\alpha \mathcal{M}_j+  \sinh\alpha \mathcal{M}^Z_j$. After rotation, we have similarly
  \begin{equation}
 \label{Malphaconjugated}
 V \mathcal{M}_j^{(\alpha)}(\gamma) V^{-1} = 
 \left(
\begin{array}{cccc}
\cosh\alpha & \cosh\alpha \sqrt{1-\gamma^2} X_j & -i \cosh\alpha Y_j & - \cosh\alpha \sqrt{1-\gamma^2} Z_j  \\
0 & \gamma \cosh\alpha Z_j  & 0 & \gamma \cosh\alpha  X_j \\
i \gamma \sinh\alpha  Y_j  & 0  &  \gamma \sinh\alpha  & 0 \\
- \sinh\alpha \sqrt{1-\gamma^2}  & - \sinh\alpha X_j & i\sinh\alpha \sqrt{1-\gamma^2}  Y_j &  \sinh\alpha Z_j  
\end{array} 
  \right) \,.
 \end{equation}

For any three consecutive sites $j,j+1,j+2$, we then have 
\begin{equation}
\label{TellT}
T(\gamma) \ell(j) T(\gamma)^\dagger = 
\mathrm{Tr}_{\mathcal{A}\otimes \mathcal{A}} ( \mathcal{M}_{L}(\gamma) \ldots \mathcal{M}^\ell_{j,j+1,j+2}(\gamma)   \ldots  \mathcal{M}_1(\gamma) ) \,,
\end{equation} 
where 
\begin{align}
\mathcal{M}^\ell_{j,j+1,j+2}(\gamma) 
\equiv 
\frac{1}{1+\gamma^2}
\left(
\mathcal{M}_{j+2} \mathcal{M}^Z_{j+1} \mathcal{M}_{j}
+
\gamma (\mathcal{M}^X_{j+2} \mathcal{M}^X_{j+1} \mathcal{M}_{j} 
+
\mathcal{M}_{j+2} \mathcal{M}^X_{j+1} \mathcal{M}^X_{j})
-
\gamma^2 \mathcal{M}_{j+2}^X \mathcal{M}^Z_{j+1} \mathcal{M}^X_{j}
\right) \,,
\end{align}
which can be brought to the following form after rotation in ancillary space:
\begin{align}
\label{Mellrotated}
V \mathcal{M}^\ell_{j,j+1,j+2}(\gamma)  V^{-1}
=\left( 
\begin{array}{cccc}
\tilde{\ell}(j) & \ldots & \ldots & \ldots \\
0 & \gamma^3  Z_j Z_{j+1}Z_{j+2}   \tilde{\ell}(j)  & 0 & \ldots \\
0 & 0 & 0 & 0 \\
0 & 0 & 0 & 0 
\end{array} \right) \,.
\end{align}
Here we have defined 
\begin{align}
 \tilde{\ell}(j) \equiv \frac{1}{1+\gamma^2}\left(Z_{j+2}+\gamma (X_{j+1} X_{j+2} +Y_{j+1} Y_{j+2}) +\gamma^2  Z_{j+1}\right)\,,
 \end{align}
 and the $\dots$ denote other combinations of the Pauli matrices which we will not need to consider.
 Indeed, from the triangular structure of \eqref{Mellrotated}, we see again that only the two non-zero diagonal entries give a non-zero contribution to the trace \eqref{TellT}. As a result we find 
\begin{align}
T(\gamma) \ell(j) T(\gamma)^{\dagger}  =  (1+\gamma^L \mathcal{Z}) \tilde{\ell}(j) \,,
\end{align} 
or, equivalently,
\begin{align}
T(\gamma) \ell(j) T(\gamma)^{-1}  = \tilde{\ell}(j) \,.
\end{align} 

The modified jump operators $\tilde{\ell}(j)$ all square to one, and commute with the global charge $Q_0 = \sum_j Z_j$. Since furthermore $T(\gamma ) H T(\gamma)^{-1} = H$, we can readily conclude that all powers of $Q_0$, or equivalently all exponentials of the form $e^{\alpha Q_0}$, are (un-normalized) NESS of the Lindbladian defined from the Hamiltonian $H$ and the jump operators $\tilde{\ell}(j)$. These form a basis for a $L+1$-dimensional space, including the identity. 

Undoing the similarity transformation, this shows that the matrices $T(\gamma)^{-1} e^{\alpha Q_0} T(\gamma)$ are (un-normalized) NESS for the Lindbladian constructed out of the Hamiltonian $H$ and jump operators $\ell(j)$.
Since $\mathcal{Z}$ commutes with both the Hamiltonian and jump operators, we can further replace $T(\gamma)^{-1}$ by $T(\gamma)^\dagger$, and conclude that the density matrices (14) in the main text are a family of NESS.

\section{Mean values in NESS}

In this Section we compute mean values of local observables in states of the form $\rho_\gamma(\beta) =
T(\gamma)^\dagger e^{\beta Q_0} T(\gamma)$. In particular, we derive eqs. (18) and (21) of the main text.

\subsection{Expectation values of matrix product operators}

We start by computing the following objects
\begin{align}
\mathcal{G}(\alpha,\beta) &= \mathrm{Tr} (e^{\alpha Q_0} T(\gamma)^{\dagger} e^{\beta Q_0} T(\gamma))  \\
\widetilde{\mathcal{G}}(\alpha,\beta) &= \mathrm{Tr} (e^{\alpha Q_0} T(\gamma)^{-1} e^{\beta Q_0} T(\gamma)) \,,
\end{align}
in terms of which we will see that all quantities of interest can be expressed.

Let us start with $\mathcal{G}(\alpha,\beta)$. Using the MPO formalism above, we can rewrite: 
\begin{align}
\mathcal{G}(\alpha,\beta) = \mathrm{Tr}_{\mathcal{A} \otimes \mathcal{A}} \prod_{j=L}^1  \mathrm{tr}_j \left(  {\mathcal{M}}_j^{(\alpha,\beta)}(\gamma)    \right) \,,
\end{align}
where ${\mathcal{M}}_j^{(\alpha,\beta)}(\gamma) \equiv {\mathcal{M}}_j^{(\alpha)}(\gamma) e^{\beta Z_j} $.
After rotation in ancillary space (see eq. \eqref{Malphaconjugated}), we have
  \begin{equation}
 \label{Malphaexpbeta}
\mathrm{tr}_j\left( V \mathcal{M}_j^{(\alpha,\beta)}(\gamma)  V^{-1} \right) = 
\left(
\begin{array}{cccc}
 2 \cosh \alpha  \cosh \beta  & 0 & 0 & -2 \sqrt{1-\gamma ^2} \cosh \alpha  \sinh \beta  \\
 0 & 2 \gamma  \cosh \alpha  \sinh \beta  & 0 & 0 \\
 0 & 0 & 2 \gamma  \sinh \alpha  \cosh \beta  & 0 \\
 -2 \sqrt{1-\gamma ^2} \sinh \alpha  \cosh \beta  & 0 & 0 & 2 \sinh \alpha  \sinh \beta  \\
\end{array}
\right) \,.
 \end{equation}
 
 The computation of $\mathcal{G}(\alpha,\beta)$ can be performed by diagonalizing \eqref{Malphaexpbeta} in ancillary space, leading to : 
 \begin{equation}
 \mathcal{G}(\alpha,\beta) =   (\lambda_1(\alpha,\beta))^L + (\lambda_2(\alpha,\beta))^L+ (\lambda_3(\alpha,\beta))^L+ (\lambda_4(\alpha,\beta))^L \,,  
 \end{equation} 
 where 
\begin{align}
\lambda_1(\alpha,\beta) &= \cosh (\alpha +\beta )+\sqrt{\cosh ^2(\alpha +\beta )-\gamma ^2 \sinh (2 \alpha ) \sinh (2 \beta )}
   \cr 
\lambda_2(\alpha,\beta) &= \cosh (\alpha +\beta )-\sqrt{\cosh ^2(\alpha +\beta )-\gamma ^2 \sinh (2 \alpha ) \sinh (2 \beta )}
   \cr
\lambda_3(\alpha,\beta) &= 2 \gamma  \sinh \alpha  \cosh \beta 
   \cr
\lambda_4(\alpha,\beta) &= 2 \gamma  \cosh \alpha  \sinh \beta 
\label{lambdaeigenvalues}
\end{align} 
are the eigenvalues of \eqref{Malphaexpbeta}.
For later use, we evaluate the function $\mathcal{G}(\alpha,\beta)$ and its derivatives at particular points : 
\begin{align}
\mathcal{G}(0,\beta) &= 2^L(\cosh \beta ^L + (\gamma \sinh\beta)^L)  \cr 
\frac{1}{ L}\partial_{\alpha}\mathcal{G}(\alpha,\beta)|_{\alpha=0} &= \left(1-\gamma ^2\right) 2^L  \tanh \beta  \cosh ^L(\beta ) \cr
\mathcal{G}(-i \pi/2,\beta) &= (-2 i)^L \left(\gamma ^L \cosh ^L\beta +\sinh ^L\beta \right)  \cr
\frac{1}{L}\partial_{\alpha}\mathcal{G}(\alpha,\beta)|_{\alpha=-i \pi/2} &= \left(1-\gamma ^2\right) (-2 i)^L  \coth \beta  \sinh ^L\beta  
\label{eigenvaluesparticularpoints}
\end{align}

We now turn to $\widetilde{\mathcal{G}}(\alpha,\beta)$. Using \eqref{Tgammam1}, we have 
\begin{align}
\widetilde{\mathcal{G}}(\alpha,\beta) &=\mathrm{Tr} (e^{\alpha Q_0} \frac{1-\gamma^L \mathcal{Z}}{1-\gamma^{2L}} T(\gamma)^{\dagger}  e^{\beta Q_0} T(\gamma))
=
\frac{\mathcal{G}(\alpha,\beta)-(i\gamma)^L \mathcal{G}(\alpha-i \pi/2,\beta)}{1-\gamma^{2L}} \,,
\label{expressionG}
\end{align}
where in the last equality we have used the identity $Z_j e^{ \alpha Z_j} = i e^{(\alpha-i \pi/2)Z_j}$.

\subsection{Mean values in NESS}

Now we compute the mean values of local operators $Z_j$ and $X_j X_{j+1} - Y_j Y_{j+1}$.
Using translation invariance, we find 
\begin{align}
\langle Z_j \rangle_\beta  \equiv \frac{\mathrm{Tr}(Z_j T(\gamma)^\dagger e^{\beta Q_0} T(\gamma)  )}{\mathrm{Tr}( T(\gamma)^\dagger e^{\beta Q_0} T(\gamma)  )} 
= \frac{\frac{1}{L}\partial_{\alpha}\mathcal{G}(\alpha,\beta)|_{\alpha=0}}{\mathcal{G}(0,\beta)} =  \frac{(1-\gamma^2)\tanh\beta}{1+ (\gamma \tanh \beta)^L} \,,
\end{align}
hence recovering eq. (18) in main text, once done the substitution $\beta \to \alpha$.

Expectation of other local operators such as $X_j X_{j+1} - Y_j Y_{j+1}$ could be similarly obtained from suitably defined generating functions, but we here compute directly :   
\begin{align}
\mathrm{Tr}(X_j X_{j+1} T(\gamma)^\dagger e^{\beta Q_0} T(\gamma)  )
= \mathrm{Tr} \left( \mathrm{Tr}_{\mathcal{A}\otimes \mathcal{A}} ( \mathcal{M}^{(0,\beta)}_{L}(\gamma) \ldots \mathcal{M}^{(X,\beta)}_{j+1}(\gamma)\mathcal{M}^{(X,\beta)}_{j}(\gamma)  \ldots  \mathcal{M}^{(0,\beta)}_1(\gamma) ) \right)\,,
\end{align} 
where $\mathcal{M}^{(X,\beta)}_{j} = X_j \mathcal{M}^{(0,\beta)}_{j} = \mathcal{M}^{X}_{j} e^{\beta Z_j}$. A similar formula holds with $X\to Y$.

Using
\begin{align}
\mathrm{tr}_j (V \mathcal{M}^{(0,\beta)}_{j} V^{-1}) &=
\left(
\begin{array}{cccc}
 2   \cosh \beta  & 0 & 0 & -2 \sqrt{1-\gamma ^2} \sinh \beta  \\
 0 & 2 \gamma   \sinh \beta  & 0 & 0 \\
 0 & 0 & 0  & 0 \\
 0  & 0 & 0 & 0 \\
\end{array}
\right)  \,, 
\\
\mathrm{tr}_j (V \mathcal{M}^{(X,\beta)}_{j} V^{-1}) &=
 \left(
\begin{array}{cccc}
0 & 2\gamma \sinh\beta  & 0 & 0 \\
2\cosh\beta & 0 & 0 & - 2\sqrt{1-\gamma^2} \sinh\beta  \\
0 & 0 & 0 & 0 \\
0 & 0 & 0 & 0 
\end{array} 
  \right) \,, 
\\
\mathrm{tr}_j (V \mathcal{M}^{(Y,\beta)}_{j} V^{-1}) &=
 \left(
\begin{array}{cccc}
0 & 0 & 0 & 0 \\
0 & 0 & 0 & 0 \\
2 i\sqrt{1-\gamma^2} \cosh\beta  & 0 &  0 & -2i \sinh\beta  \\
 0  & 0  & -2i \gamma  \cosh\beta & 0 \\
\end{array} 
  \right) \,,
\end{align}
we find 
\begin{align}
\langle X_j X_{j+1}-Y_j Y_{j+1} \rangle_\beta  = \frac{\mathrm{Tr}((X_j X_{j+1}-Y_j Y_{j+1}) T(\gamma)^\dagger e^{\beta Q_0} T(\gamma)  )}{\mathrm{Tr}( T(\gamma)^\dagger e^{\beta Q_0} T(\gamma)  )} 
 = \frac{\gamma  \left(2-\gamma ^2\right) \tanh (\beta )+(\gamma  \tanh (\beta ))^{L-1}}{\gamma ^L \tanh ^L(\beta )+1} \,,
\end{align}
which recovers (21) in main text.

\section{Late time expectation values -- Gibbs ensemble} 

Here we consider real time evolution in the Linbdlad system, and the emergence of the Gibbs Ensemble. We show how to
compute the Lagrange multiplier of the Gibbs ensemble, which eventually leads to a prediction for the long time limit of
local observables. The computations in this Section are performed in the infinite volume limit. Exact finite volume
computations confirming the general statements for specific cases will be provided in the next Section.

The conservation of $Q_\gamma$ implies that if a Gibbs ensemble $\rho_G\sim e^{-\lambda Q_{\gamma}}$ emerges during time evolution, then it has to satisfy
\begin{equation}
  \text{Tr}\Big(\rho_0Q_\gamma\Big)=\frac{  \text{Tr}\Big(e^{-\lambda Q_\gamma}Q_\gamma\Big)}{  \text{Tr}e^{-\lambda Q_\gamma}}.
\end{equation}
This equation can be used to fix $\lambda$ using knowledge of the initial state.

Once $\lambda$ is found, the predictions for the steady state values of observables can be given by results from the
previus subsection. For example, for the mean value of $Z_j$ we find
\begin{equation}
  \vev{Z_j}=\frac{  \text{Tr}\Big(e^{-\lambda Q_\gamma}Q_\gamma\Big)}{  \text{Tr}e^{-\lambda Q_\gamma}}=-(1-\gamma^2)\tanh(\lambda).
\end{equation}

Using the similarity transformation the r.h.s. is actually found to be
\begin{equation}
 -\tanh(\lambda)
\end{equation}
Furthermore, the l.h.s. can be expressed as
\begin{equation}
   \text{Tr}\Big(T(\gamma)\rho_0T^\dagger(\gamma)Q_0\Big)
\end{equation}

If the initial density matrix is chosen to be
\begin{equation}
  \rho_0\sim e^{\beta Q_0}
\end{equation}
then once again we can use the results from the next subsection to conclude
\begin{equation}
  (1-\gamma^2)\tanh(\beta)=-\tanh(\lambda).
\end{equation}
Combining everything we obtain the prediction for the long time limit
\begin{equation}
  \vev{Z_j}=(1-\gamma^2)^2 \tanh(\beta)
\end{equation}

\section{Late time expectation values -- exact computations} 

In this section we detail the computation of the late-time expectation values of local observables following a quantum
quench from initial states given by eq. (23) in main text. In particular we focus on the local observable $\langle Z_j
\rangle$.

Let us start by recalling that an over-complete basis of the $L+1$-dimensional space of NESS can be generated by the $\tilde{\rho}_\gamma(\alpha) = T(\gamma)^{-1} e^{\alpha Q_0} T(\gamma)$. 
Since $Q_0 = \sum_j Z_j$ has $L+1$ distinct eigenvalues of the form $2n-L$ with $n=0,\ldots, L$, we can alternatively define a basis of the space of NESS in terms of the projectors $\widetilde{P}_n = T(\gamma)^{-1} P_n T(\gamma)$, where 
\begin{equation}
P_n = \frac{1}{L+1} \sum_{k=0}^L e^{i\frac{2\pi k}{L+1}(\frac{L+Q_0}{2}-n)} 
\label{eq:Pndef}
\end{equation}
is the projector onto the subspace where $Q_0$ has eigenvalue $2n-L$.

As a consequence, any density matrix in the space of NESS can be decomposed as : 
\begin{equation}
\rho_{\rm NESS} = \sum_{n=0}^L \frac{\mathrm{Tr}(\rho_{\rm NESS}\widetilde{P}_n)}{\mathrm{Tr}(\widetilde{P}_n)}  \widetilde{P}_n.
\end{equation} 
Starting from an arbitrary $\rho(t=0)$, we therefore have at late times 
\begin{equation}
\lim_{t\to\infty} e^{\mathcal{L}t} \rho(t=0) =  \sum_{n=0}^L \frac{\mathrm{Tr}(\rho(t=0)\widetilde{P}_n)}{\mathrm{Tr}(\widetilde{P}_n)}  \widetilde{P}_n
\end{equation}
and therefore, for any observable $\mathcal{O}$, 
\begin{equation}
\lim_{t\to\infty} \langle \mathcal{O} \rangle =  \sum_{n=0}^L \frac{\mathrm{Tr}(\rho(t=0)\widetilde{P}_n)}{\mathrm{Tr}(\widetilde{P}_n)}  \mathrm{Tr}(\widetilde{P}_n \mathcal{O}) \,.
\label{eq:limO}
\end{equation}

All the traces involved in \eqref{eq:limO} can be computed using the matrix product operator techniques. 

Focusing on the observable $Z_j$ we need to compute : 
\begin{equation}
\lim_{t\to\infty} \langle Z_j \rangle =  \sum_{n=0}^L \frac{\mathrm{Tr}(\rho(t=0)\widetilde{P}_n)}{\mathrm{Tr}(\widetilde{P}_n)}  \mathrm{Tr}(\widetilde{P}_n Z_j) \,.
\label{eq:limZ}
\end{equation}

The first trace $\mathrm{Tr}(\widetilde{P}_n) = \mathrm{Tr}(P_n)$ can easily be computed without resorting to MPO techniques, as it corresponds to the dimension of the eigenspace of $P_n$ with eigenvalue $2L-n$, however as a warm-up we present its computation using the previously computed function $\widetilde{\mathcal{G}}(\alpha,\beta)$. Using the decomposition \eqref{eq:Pndef} of the projector $P_n$, we have 
\begin{align}
 \mathrm{Tr}(\widetilde{P}_n)  
 &= 
 \frac{1}{L+1} \sum_{k=0}^L 
e^{i\frac{2\pi k}{L+1}(\frac{L}{2}-n)}  
 \mathrm{Tr} \left( T(\gamma)^{-1} e^{i \frac{\pi k}{L+1} Q_0} T(\gamma) \right)
 \cr 
&=  \frac{1}{L+1} \sum_{k=0}^L 
e^{i\frac{2\pi k}{L+1}(\frac{L}{2}-n)}  
  \widetilde{\mathcal{G}}(0, \frac{i k\pi }{L+1})
\cr 
&=  \frac{{2^L}}{L+1} \sum_{k=0}^L 
e^{i\frac{2\pi k}{L+1}(\frac{L}{2}-n)} (\cos \frac{k\pi}{L+1})^L 
\cr 
&=  {L \choose n}\,.
\label{trP}
\end{align}

The second trace can be similarly evaluated as :
\begin{align}
 \mathrm{Tr}(\widetilde{P}_n Z_j)  
&=  \frac{1}{L+1} \sum_{k=0}^L 
e^{i\frac{2\pi k}{L+1}(\frac{L}{2}-n)}  
\frac{1}{L}  \partial_{\alpha} \widetilde{\mathcal{G}}(\alpha, \frac{i k\pi }{L{+1}})|_{\alpha=0}
\cr 
&=  i 2^L \frac{1-\gamma^2}{1-\gamma^{2L}}\frac{1}{L+1} \sum_{k=0}^L 
e^{i\frac{2\pi k}{L}(\frac{L}{2}-n)} (\sin\frac{k\pi}{L+1}(\cos\frac{k\pi}{L+1})^{L-1}+ (i \gamma)^L \cos\frac{k\pi}{L+1}(\sin\frac{k\pi}{L+1})^{L-1}) 
\cr 
&= - \frac{1-\gamma^2}{1-\gamma^{2L}} (1-(-1)^{L-n} \gamma^L )  \frac{L-2n}{n}  
{L-1 \choose n-1}\,.
\label{TrPnZj}
\end{align}

We now move to the third trace, $ \mathrm{Tr}(\rho(t=0) \widetilde{P}_n )$. Taking the normalized density matrix $\rho(t=0) = e^{ \alpha Q_0} / (2\cosh \alpha)^L $, 
\begin{align}
 \mathrm{Tr}(\rho(t=0) \widetilde{P}_n )  
&=   \frac{1}{(2\cosh\alpha)^L}\frac{1}{L+1} \sum_{k=0}^L 
e^{i\frac{2\pi k}{L+1}(\frac{L}{2}-n)}  
\widetilde{\mathcal{G}}(\alpha, \frac{i k\pi }{L+1})   
\cr 
&=
   \frac{1}{(2\cosh\alpha)^L}\frac{1}{L+1} \frac{1}{1-\gamma^{2L}}
   \left(  \mathcal{F}_n(\alpha) - (i\gamma)^L \mathcal{F}_n(\alpha-i \pi/2)  \right)   
   \label{Trrho0Pntilde}
\end{align}
where in the last line we have used the expression \eqref{expressionG} of $\widetilde{\mathcal{G}}$ in terms of $\mathcal{G}$, and introduced the functions
\begin{align}
\mathcal{F}_n(\alpha) &\equiv 
 \sum_{k=0}^L 
e^{i\frac{2\pi k}{L+1}(\frac{L}{2}-n)}  
\widetilde{\mathcal{G}}(\alpha, \frac{i k\pi }{L+1})
\\ 
&=  \sum_{k=0}^L 
e^{i\frac{2\pi k}{L+1}(\frac{L}{2}-n)}  
\left( 
\lambda_1(\alpha, \frac{i k\pi }{L+1})^L 
+
\lambda_2(\alpha, \frac{i k\pi }{L+1})^L 
+
\lambda_3(\alpha, \frac{i k\pi }{L+1})^L 
+\lambda_4(\alpha, \frac{i k\pi }{L+1})^L 
\right) 
\\ 
&\equiv 
\mathcal{F}^{(1)}_n(\alpha) 
+ 
\mathcal{F}^{(2)}_n(\alpha)
+ 
\mathcal{F}^{(3)}_n(\alpha)
+
\mathcal{F}^{(4)}_n(\alpha)
\end{align}

Using the expressions \eqref{lambdaeigenvalues} of the eigenvalues $\lambda_i$, the contributions $\mathcal{F}^{(3)}$ and $\mathcal{F}^{(4)}$ are easily evaluated. We find : 
 \begin{align} 
  \frac{1}{({2}\cosh\alpha)^L}\frac{1}{L+1} \frac{1}{1-\gamma^{2L}} \mathcal{F}_n^{(3)}(\alpha) &=\frac{1}{{2^L}}\frac{( \gamma  \tanh \alpha)^L }{1-\gamma^{2L}} \left(
\begin{array}{c}
 L \\
 n \\
\end{array}
\right) \,,
\label{F3}
\\
  \frac{1}{({2}\cosh\alpha)^L}\frac{1}{L+1} \frac{1}{1-\gamma^{2L}} \mathcal{F}_n^{(4)}(\alpha) &=\frac{1}{{2^L}} \frac{\gamma^L}{1-\gamma^{2L}}(-1)^{L-n} \left(
\begin{array}{c}
 L \\
 n \\
\end{array}
\right) 
\,.
\label{F4}
\end{align}

We now move to the contribution $\mathcal{F}^{(1)}+\mathcal{F}^{(2)}$. Using the expression \eqref{lambdaeigenvalues}, 
\begin{align}
\lambda_1(\alpha,\beta)^L + \lambda_2(\alpha,\beta)^L
= 2 \sum_{\substack{j=0 \\ j~{\rm even}}}^L 
{L \choose j}  \left( \cosh (\alpha +\beta ) \right)^{L-j} \left(\cosh ^2(\alpha +\beta )-\gamma ^2 \sinh (2 \alpha ) \sinh (2 \beta )\right)^{j/2} \,.
\end{align}
Hence, 
\begin{align}
& \mathcal{F}^{(1)}_n(\alpha) 
+ 
\mathcal{F}^{(2)}_n(\alpha) = 
\cr 
&
\frac{2}{2^L} \sum_{k=0}^L 
  \sum_{\substack{j=0 \\ j~{\rm even}}}^L 
{L \choose j}  e^{- \alpha L } e^{\frac{- 2 i n k \pi}{L+1}}  
 \left(1+e^{\frac{2 i k \pi}{L+1}} e^{2\alpha} \right)^{L-j} \left((1+e^{\frac{2 i k \pi}{L+1}} e^{2\alpha})^2 -  \gamma^2  (1-e^{\frac{4 i k \pi}{L+1}})(1-e^{4\alpha}) \right)^{j/2}
 \\ 
&= 
 \frac{e^{- \alpha L }}{2^{L-1}} \sum_{k=0}^L 
  \sum_{l=0}^{L/2} 
  \sum_{a=0}^{L-2l} 
    \sum_{\substack{b_1, b_2 \geq 0  \\ b_1+b_2 \leq l} }
{L \choose 2l}{L-2l \choose a} {l \choose b_1,b_2,l-b_1-b_2} 
 e^{\frac{2 i (a+b_1 + 2 b_2-n) k \pi}{L+1}}  
e^{2 a \alpha}  (2  e^{2 \alpha}  )^{b_1} \frac{(e^{4\alpha}(1-\gamma^2)+\gamma^2)^{b_2} }{(1-\gamma^2 (1-e^{4\alpha}))^{b_1+b_2-l}}
\cr
&= 
 \frac{L+1}{e^{ \alpha L }}\frac{2}{2^{L}} 
  \sum_{l=0}^{L/2} 
    \sum_{\substack{b_1, b_2 \geq 0  \\ b_1+b_2 \leq l} }
    \frac{L!\,e^{2 n \alpha} 
(1-\gamma^2 (1-e^{4\alpha}))^{l-b_1} 2^{b_1} }{(2l)! (n-b_1-2b_2)! {(L-2l-n+b_1+2b_2)!}}
 {l \choose b_1,b_2,l-b_1-b_2}   
\left(\frac{1-\gamma^2 (1-e^{-4\alpha})}{1-\gamma^2 (1-e^{4\alpha})}\right)^{b_2} \,.
\cr 
\end{align}

We further expand this expression to write it as a polynomial in $\gamma$ (the deformation parameter) and we obtain (after proper re-arranging the sum)

\begin{align}
\frac{(L+1) \,L!\, e^{2 \alpha  n}}{2^{L-1} e^{\alpha  L}} \sum\frac{l!\, 2^{l-b_3} \left(e^{-4 \alpha }-1\right)^t \left(e^{4 \alpha }-1\right)^{f-t}}{(2 l)!\, t!\, (b_2-t)! (l-b_3)! (f-t)! (t-b_2+b_3-f)! (n-2 b_2+b_3-l)! (2 b_2-b_3-l+L-n)!}\gamma ^{2 f} ,
\end{align}
where we used the shortcut 
\begin{align}
&\sum\to \sum_{f=0}^{L/2} \sum_{l=f}^{L/2} \sum_{b_3=f}^l \sum_{b_2=0}^{b_3} \sum_{t=0}^{b_2}.
\end{align}
We used the software Mathematica 12.3 to further simplify this expression and we obtained
\begin{align}
\mathcal{F}^{(1)}_n(\alpha) 
+ 
\mathcal{F}^{(2)}_n(\alpha) = & \kappa\sum_{f=0}^{L/2} \frac{  \left(e^{4  \alpha }-1\right)^f \gamma ^{2 f}    (L-f)!  }{f! n! \,e^{ \alpha  (L-2 n)}\,2^{L-n}} {_3\tilde{F}}_2\left(-f,\frac{1-n}{2},-\frac{n}{2};\frac{ L-n-2 f+1}{2},\frac{L-n-2 f+2}{2};\frac{1}{e^{4  \alpha }}\right),
\label{F12}
\end{align}
where $\kappa=(L+1)\sqrt{\pi } L$ and $_3\tilde{F}_2$ is the hypergeometric function regularized.

Reporting the results \eqref{F3}, \eqref{F4} and \eqref{F12} into \eqref{Trrho0Pntilde} (where the terms $\mathcal{F}_n (\alpha-i \pi/2)$ can just be obtained by shifting the argument), we get :

\begin{align}
\text{Tr} (\rho(t=0)\tilde P_n)=&\frac{1}{ {2^L}}\frac{(-1)^n \gamma ^L }{\gamma ^{2 L}-1}\binom{L}{n} \left((-\gamma  \tanh (\alpha ))^L-(-1)^{n} \tanh ^L(\alpha )+(-1)^n \gamma ^L-(-1)^{L}\right)+\nonumber\\
&\frac{1}{( {2}\cosh \alpha)^L} \frac{1}{1-\gamma^{2L}}\sum_{f=0}^{L/2} \frac{\sqrt{\pi } L  \left(1-e^{4  \alpha }\right)^f \gamma ^{2 f} (-1)^{f+n} 2^{n-L}  (L-f)! e^{- \alpha  (L-2 n)} \left((-1)^{L+1} \gamma ^L+(-1)^n\right)}{f! n!}\nonumber\\
&\, {_3\tilde{F}}_2\left(-f,\frac{1-n}{2},-\frac{n}{2};\frac{ L-n-2 f+1}{2},\frac{L-n-2 f+2}{2};\frac{1}{e^{4  \alpha }}\right).
\label{Trrho0Pn}
\end{align}

The three factors \eqref{trP}, \eqref{TrPnZj} and \eqref{Trrho0Pn} can now be gathered in the initial expression \eqref{eq:limZ} for $\lim \langle Z_j\rangle$. Performing the sum over $n$, we find that the contributions coming from $\mathcal{F}^{(3)}_n$ and $\mathcal{F}^{(4)}_n$ vanish.  
It remains to compute
\begin{align}
 \lim_{t\to \infty}\langle Z_j \rangle=&\frac{\sqrt{\pi } \left(1-\gamma ^2\right) e^{  -\alpha L} }{{2^L}\left(\gamma ^{2 L}-1\right)^2(\sinh (2 \alpha ) \text{csch}(\alpha ))^{L}}\sum_{n=0}^L \sum_{f=0}^{L/2}\frac{ 2^n \left(e^{4 \alpha }-1\right)^f \gamma ^{2 f} (2 n-L) e^{2 \alpha  n}  (L-f)! \left((-1)^n-(-\gamma)^{L}\right)^2}{ (f+1)!  (n+1)!} \nonumber\\
&\, {_3\tilde{F}}_2\left(-f,\frac{1-n}{2},-\frac{n}{2};\frac{ L-n-2 f+1}{2},\frac{L-n-2 f+2}{2};\frac{1}{e^{4  \alpha }}\right) \,.
\label{almostfinalZ} 
\end{align}
This expression looks complicated at first sight, but we will now see that it is equivalent to the expression (27) in the main text. Both expressions contain a prefactor $\frac{1-\gamma^2}{(1-\gamma^{2L})^2}$ which we therefore omit in the following, and compare the remaining polynomials in $\gamma$ order by order. Starting from (27), the remaining polynomial takes the form 
\begin{align}
\left(1-\gamma ^2\right) \tanh (\alpha ) \left(1+\gamma ^{2 L}-2 \gamma ^L \tanh ^{L-2}\alpha \right),
\label{Mariusnodenominator}
\end{align}
in particular the exponents of $\gamma$ that gives a contribution different from 0 are $0, 2, L, 2L, L+2, 2L+2$. We shall now demonstrate that all other powers indeed vanish in the polynomial associated with expression \eqref{almostfinalZ}.
In \eqref{almostfinalZ}, the coefficients are $2f, 2f+L, 2f+2L$, so the only non-zero contribution should come from $f=0,1,L/2$, the last one only for $L$ even.

By direct computation, we found that to obtain \eqref{Mariusnodenominator} it is enough to sum the contribution of $f=0$ and $f=1$. Let us show that the other terms vanish. 
First, we consider the contribution of $f=L/2$ in \eqref{almostfinalZ}. This is proportional to
\begin{align}
\sum_{n=0}^L \frac{(L-2 n) e^{2 \alpha  n} \left((-1)^n-(-\gamma)^{L}\right)^2\sin (\pi  n)}{n } \, \, _2F_1\left(-\frac{L}{2},-\frac{n}{2};1-\frac{n}{2};\frac{1}{e^{4 \alpha }}\right)\label{F21}.
\end{align}
Since the function $ _2F_1\left(-\frac{L}{2},-\frac{n}{2};1-\frac{n}{2};e^{-4 \alpha }\right)$ is finite for $n$ odd, \eqref{F21} vanishes due to $\sin (n \pi)$,  while for $n$ even:
\begin{align}
\, _2F_1\left(-\frac{L}{2},-\frac{n}{2};1-\frac{n}{2};x\right)=\sum_{k=0}^{L/2} \frac{(-1)^k n \binom{L/2}{k} x^k}{n-2 k},
\end{align}
and if we substitute this into \eqref{F21}, we get
\begin{align}
\sum_{k=0}^{L/2}\sum_{n=0}^L \frac{(-1)^k (L-2 n) \sin (\pi  n) \binom{\frac{L}{2}}{k} e^{2 \alpha  (n-2 k)}}{n-2 k}=\sum_{n=0}^L i^n (L-2 n) \binom{\frac{L}{2}}{\frac{n}{2}}=0,
\end{align}
where since $\sin (n \pi)$ is zero, we need to only keep the singular term.

It remains to show that all the terms with $f>1$ do not contribute.  
Removing the irrelevant terms and considering $e^{-\alpha}=z$, we should prove that the following term vanishes
\begin{align}
\sum_{n=0}^{L}\frac{(L-2 n) z^{L-2 n} \, }{ (n+1)!  (1-2 f+L-n)!} {_3F_2}\left(-f,\frac{1-n}{2},-\frac{n}{2};\frac{ L-n-2 f+1}{2},\frac{L-n-2 f+2}{2};z^4\right).
\end{align}
Expanding as a series in $z$ and re-shifting the sum over $n$ we obtain
\begin{align}
\sum_{m=0}^\infty \sum_{n=-2m}^{L-2m} \frac{(-1)^{m} \binom{f}{m} (L-4 m-2 n) z^{L-2 n}}{(n+1)! \, (1-2 f+L-n)!}=\sum_{m=0}^\infty \sum_{n=0}^{L} \frac{(-1)^{m} \binom{f}{m} (L-4 m-2 n) z^{L-2 n}}{(n+1)! \, (1-2 f+L-n)!}.
\end{align}
We can now sum over $m$ and we are left with
\begin{align}
\sum_{n=0}^L (L-2 n) \, _1F_0(-f;;1)+4 f \, _1F_0(1-f;;1).
\label{lasteqhyperg}
\end{align}
Considering that
\begin{align}
_1F_0(a;;x)=(1-x)^{-a},
\end{align}
each terms of \eqref{lasteqhyperg} are zero for $f>1$ as stated at the beginning.

To summarize, we proved that \eqref{eq:limZ} is equivalent to the expression (27) given in the main text.

\bibliography{pozsi-general}